\documentclass[%
 reprint,
 amsmath,amssymb,
 aps,
]{revtex4-1}

\usepackage[dvipdfmx]{graphicx} 
\usepackage{dcolumn}
\usepackage{bm}
\usepackage[font=small,format=plain,labelformat=simple,labelsep=period,justification=raggedright]{caption}
\usepackage{amsmath}
\usepackage{here}
\usepackage{natbib}
\usepackage[dvipdfmx]{color}
\usepackage{epstopdf}
\usepackage{comment}
\usepackage{multirow}

\begin{document}

\preprint{APS/123-QED}

\title{Two-lane totally asymmetric simple exclusion process with extended Langmuir kinetics}

\author{Hiroki Yamamoto$^{1}$}
\email{h18m1140@hirosaki-u.ac.jp}
\thanks{}%
\author{Shingo Ichiki$^{2}$}%
\author{Daichi Yanagisawa$^{2,3}$}%
\author{Katsuhiro Nishinari$^{2,3}$}%
\affiliation{%
$^{1}$ School of Medicine, Hirosaki University, 5 Zaifu-cho Hirosaki city, Aomori, 036-8562, Japan\\
$^{2}$ Research Center for Advanced Science and Technology, The University of Tokyo,\\
4-6-1 Komaba, Meguro-ku, Tokyo 153-8904, Japan\\
$^{3}$ Department of Aeronautics and Astronautics, School of Engineering, The University of Tokyo,\\
7-3-1 Hongo, Bunkyo-ku, Tokyo 113-8656, Japan
}%

\date{\today}

\begin{abstract}
Multi-lane totally asymmetric simple exclusion processes with interactions between the lanes have recently been investigated actively. This paper proposes a two-lane model with extended Langmuir kinetics on a periodic lattice. Both bidirectional and unidirectional flows are investigated. In our model, the hopping, attachment, and detachment rates vary depending on the state of the corresponding site in the other lane. We obtain a theoretical expression for the global density of the system in the steady state from three kinds of mean-field analyses (1-, 2-, and 4-cluster cases). We verify that the 4-cluster mean-field analysis approximates well the results of computer simulations for the two directional flows and reproduces the differences between them. We expect these findings to contribute to a deeper understanding of the dynamic features of actual traffic systems.
\end{abstract}

\maketitle

\maxdeadcycles=1000


\section{INTRODUCTION}

The asymmetric simple exclusion process (ASEP), which is a stochastic process involving particles on a lattice, has been applied in many fields~\cite{schadschneider2010stochastic} since it was first proposed by MacDonald and Gibbs~\cite{macdonald1968kinetics,macdonald1969concerning}. A special version of ASEP, in which particles on a lattice can hop unidirectionally, is referred to as a totally asymmetric simple exclusion process (TASEP). Researchers have applied TASEPs to traffic flows of self-driven particles, as in biological transport~\cite{chou2011non,appert2015intracellular,PhysRevE.99.052122,10.1371/journal.pone.0182178}, vehicular traffic~\cite{RevModPhys.73.1067,yamamoto2017velocity}, and pedestrian flow~\cite{arita2015exclusive,PhysRevE.100.042106,CDA8}. Recently, one-lane TASEPs with varying hopping probabilities~\cite{PhysRevE.82.022103,hao2016exponential} and multi-lane TASEPs with interactions between lanes~\cite{lin2011bidirectional,jiang2009phase,hao2018theoretical,PhysRevE.100.032133,ezaki2011positive,tsuzuki2018effect,PhysRevE.98.042102} have begun to be investigated. For example, in Refs.~\cite{PhysRevE.82.022103,hao2016exponential}, the hopping probability of a particle varies depending on the states of the sites surrounding it, whereas in Refs.~\cite{jiang2009phase,hao2018theoretical,PhysRevE.100.032133}, the hopping rate of a particle varies depending on the state of the other lane.

One of the extensions of TASEP, a TASEP with Langmuir kinetics, which we refer to as LK-TASEP in the present paper, has begun to be actively investigated~\cite{PhysRevLett.90.086601,PhysRevE.68.026117,PhysRevE.70.046101,WANG2008457,dhiman2014two,sharma2017phase,GARG2019123356,vuijk2015driven,ichiki2016totally,ichiki2016totally2,yanagisawa2016totally,PhysRevE.98.042119,nishinari2005intracellular,miedema2017correlation,kushwaha2020crowding}. In the original LK-TASEP, a particle attaches (detaches) at a certain rate $\omega_{\rm A}$ ($\omega_{\rm D}$) when the targeted site on the lattice is empty (occupied). The LK-TASEP was first proposed in Ref.~\cite{PhysRevLett.90.086601} and was investigated using mean-field theory in Refs.~\cite{PhysRevE.68.026117,PhysRevE.70.046101}. Recently, multi-lane LK-TASEPs have been studied~\cite{WANG2008457,dhiman2014two,sharma2017phase,GARG2019123356}. In addition, Refs.~\cite{vuijk2015driven,ichiki2016totally,ichiki2016totally2,yanagisawa2016totally,PhysRevE.98.042119} changed the attachment and detachment rates depending on the occupancy of the adjacent sites. We note that the models of Refs.~\cite{ichiki2016totally,ichiki2016totally2,yanagisawa2016totally} are more generalized versions of those discussed in Ref.~\cite{PhysRevE.68.026117}.
 
In the present paper, we consider a two-lane extended LK-TASEP on a periodic lattice. The proposed model considers the interaction between particles in each lane without lane changing. Specifically, the hopping rate $p$ and the attachment (detachment) rate $\omega_{\rm A}$ ($\omega_{\rm D}$) vary depending on the state of the corresponding site in the other lane. We stress that the hopping rule has already been employed in Refs.~\cite{jiang2009phase,hao2018theoretical,PhysRevE.100.032133}; however, the attachment and detachment rule has not been considered in previous studies. In the presnt paper, we investigate both unidirectional and bidirectional flows. We conduct computer simulations and perform three kinds of mean-field analyses (1-, 2-, and 4-cluster cases) to investigate the global density of the system in the steady state. We find that the 4-cluster mean-field analysis reproduces the results of the simulations well.

From a practical point of view, LK-TASEP has been used extensively for analyzing the motions of motor proteins in Refs.~\cite{nishinari2005intracellular,miedema2017correlation,kushwaha2020crowding}, in which the investigators determined the parameters from experimental data. 
For applications to traffic flow, a TASEP model with an absorbing lane, which can be classified into the same category as LK-TASEP, has been investigated in Refs.~\cite{ezaki2011positive} (parking lots) and \cite{tsuzuki2018effect,PhysRevE.98.042102} (airport transportation systems). 
Our proposed model can also be applied to traffic flow; e.g., to crowd dynamics in the situations where multiple lanes are formed in a narrow passage, such as in trains, airplanes, and concert halls. In those situations, we observe two important phenomena. First, we often observe that people decrease their walking speed to avoid collisions while walking side-by-side in unidirectional flows (passing each other in bidirectional flows). Second, decision making during inflow to (outflow from) a passage is influenced by the local state. For example, people tend to hesitate to enter a passage when it is congested locally at the inflow point. The former phenomenon corresponds to a change of the hopping rate, whereas the latter one corresponds to changes of attachment and detachment rates. 

This paper is organized as follows. Section \ref{sec:model} describes the details of our proposed model. In Sec. \ref{sec:mean}, the numerical results from mean-field analyses are presented. Section \ref{sec:comparison} compares the results from mean-field analyses and simulation results. Finally, the paper concludes in Sec. \ref{sec:conclusion}.

\section{MODEL}
\label{sec:model}
\begin{figure}[htbp]
\begin{center}
\includegraphics[width=8.5cm,clip]{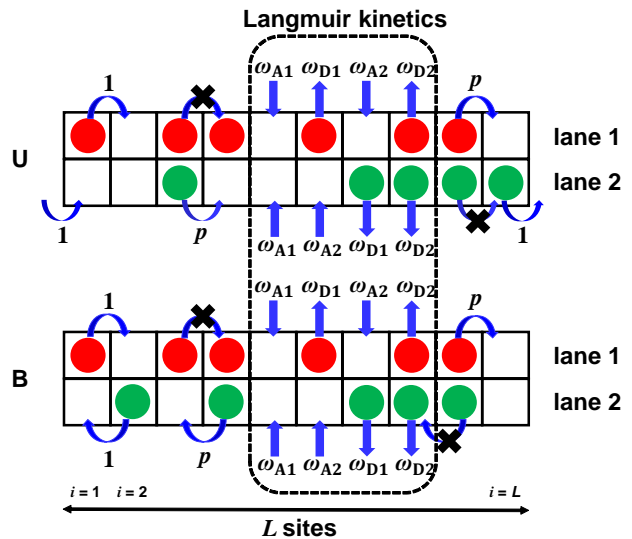}
\caption{(Color Online) Schematic illustration of the present model. The upper and lower panels represent unidirectional (U) and bidirectional (B) flows, respectively. Red (green) circles represent particles in lane 1 (2). The left and right boundaries are connected (periodic boundary conditions). We note that this figure shows the case $L=10$.}
\label{fig:model}
\end{center}
\end{figure}

The model consists of two $L$-site lanes, labeled $i=1,2,...,L$, as shown in Fig. \ref{fig:model}. Each site can either be empty or be occupied by one particle. The state of a site is represented by 1 if a particle occupies that site; otherwise, its state is represented by 0. We employ periodic boundary conditions, i.e., site $L$ and site 1 are connected, and we use random updating. Changing lanes is prohibited in this model. In the present paper, we consider two cases: unidirectional flows (particles in both lanes all hop in the same direction) and bidirectional flows (the particles in the two lanes hop in opposite directions). 

Next, we describe the update scheme for the case of a unidirectional flow. In the proposed model, the hopping rate and attachment (detachment) rates depend on the occupancy of the corresponding site in the other lane. 

For the hopping rate, a particle at site $i$ in one lane hops to site $(i+1)$ with rate $1$ if site $i$ in the other lane is vacant; otherwise, it hops with rate $p$ ($0\leq p\leq1$). 

However, for the attachment and detachment rates, the process can be divided into two patterns; specifically, 
(i) if site $i$ in one lane is empty and site $i$ in the other lane is empty (occupied), the particle attaches with rate $\omega_{\rm A1}$ ($\omega_{\rm A2}$), whereas
(ii) if site $i$ in one lane is occupied and site $i$ in the other lane is empty (occupied), the particle detaches with rate $\omega_{\rm D1}$ ($\omega_{\rm D2}$). For bidirectional flows, only the hopping rule for lane 2 differs from the case of unidirectional flows. In this case, a particle at site $i$ in lane 2 hops to site ($i-1$) with rate $1$ if the corresponding site in lane1 is vacant; otherwise, it hops with rate $p$ ($0\leq p\leq1$). We note that this model yields the standard LK-TASEP when $p=1$, $\omega_{\rm A1}=\omega_{\rm A2}$, and $\omega_{\rm D1}=\omega_{\rm D2}$.

\section{MEAN-FIELD ANALYSES}
\label{sec:mean}
In this section, we investigate the density profile in the steady state using three kinds of mean-field analyses. We hereafter write the probability finding configuration $\tau$ as $P(\tau)$. The configuration $\tau$, which is one element of the set $S$ which consists of all possible configurations, contains ($2\times L$) figures. The top (bottom) row represents the state of the sites in lane 1 (2). For example, when $L=5$ and the occupied site numbers in lane 1 are 1, 2, and 3, and those in lane 2 are 4 and 5, $\tau$ can be written as
\begin{equation}
\tau=\begin{smallmatrix}  1  & 1 & 1 & 0 & 0 \\0 & 0 & 0 & 1 & 1 \end{smallmatrix}.
\end{equation}

The master equation for this system can be written as
\begin{equation}
\frac{d P(\tau)}{dt}=\sum_{\tau^{\prime}\in S} W(\tau^{\prime}\to\tau)P(\tau^{\prime})-\sum_{\tau\in S} W(\tau\to\tau^{\prime})P(\tau),
\end{equation}
where $W(\tau^{\prime}\to\tau)$ is the transition weight to go from state $\tau^{\prime}$ to state $\tau$. However, it is very difficult to analyze ($2\times L$)-site configurations. We therefore consider cluster approximations; specifically, 1-, 2-, and 4-cluster approximations. In the following subsections, we consider both unidirectional and bidirectional flows.

\subsection{One-cluster mean-field analysis}
\label{sec:1cluster}
In this subsection, we consider the 1-cluster mean-field analysis, which is identical to the normal mean-field analysis used in ASEP investigations. Translational invariance leads to spatial homogeneity, i.e., the probability is independent of the site number; therefore, we can abbreviate the site number in the following discussions (similarly in Subsec. \ref{sec:2cluster} and \ref{sec:4cluster}).

For the probability $P(\begin{smallmatrix} 1 \\ \ast \end{smallmatrix})$, where $\ast$ represents either 0 or 1, the master equation for unidirectional flows can be written as
\begin{equation}
\begin{split}
\frac{d P(\begin{smallmatrix} 1 \\ \ast \end{smallmatrix})}{dt}=&\bigl[P(\begin{smallmatrix} 1 & 0 \\ 0  & \underline{\ast} \end{smallmatrix})+p P(\begin{smallmatrix} 1 & 0 \\ 1 & \underline{\ast} \end{smallmatrix})+\omega_{\rm A1} P(\begin{smallmatrix} 0 \\ \underline{0} \end{smallmatrix})+\omega_{\rm A2} P(\begin{smallmatrix} 0 \\ \underline{1} \end{smallmatrix})\bigr]\\
&-\bigl[P(\begin{smallmatrix} 1 & 0 \\ \underline{0} & \ast \end{smallmatrix})+p P(\begin{smallmatrix} 1 & 0 \\ \underline{1} & \ast \end{smallmatrix})+\omega_{\rm D1} P(\begin{smallmatrix} 1 \\ \underline{0} \end{smallmatrix})+\omega_{\rm D2} P(\begin{smallmatrix} 1 \\ \underline{1} \end{smallmatrix})\bigr],
\end{split}
\label{eq:original1}
\end{equation}
where the underlined sites in the right hand corresponds to the sites in the left hand (and similarly hereafter). 
We note that this equation does not change for bidirectional flows, although that of $P(\begin{smallmatrix} \ast \\ 1 \end{smallmatrix})$ changes.

Performing the mean-field analysis, i.e., ignoring higher correlations in Eq. (\ref{eq:original1}), we have 
\begin{equation}
\begin{split}
\frac{d P(\begin{smallmatrix} 1 \\ \ast \end{smallmatrix})}{dt}=&\bigl[P(\begin{smallmatrix} 1 \\ \ast \end{smallmatrix})P(\begin{smallmatrix} 0 \\ \ast \end{smallmatrix})P(\begin{smallmatrix} \ast \\ 0 \end{smallmatrix})+pP(\begin{smallmatrix} 1 \\ \ast \end{smallmatrix})P(\begin{smallmatrix} 0 \\ \ast \end{smallmatrix})P(\begin{smallmatrix} \ast \\ 1 \end{smallmatrix})\\
&+\omega_{\rm A1}P(\begin{smallmatrix} 0 \\ \ast \end{smallmatrix})P(\begin{smallmatrix} \ast \\ 0 \end{smallmatrix})+\omega_{\rm A2}P(\begin{smallmatrix} 0 \\ \ast \end{smallmatrix})P(\begin{smallmatrix} \ast \\ 1 \end{smallmatrix})\bigr]\\
&-\bigl[P(\begin{smallmatrix} 1 \\ \ast \end{smallmatrix})P(\begin{smallmatrix} 0 \\ \ast \end{smallmatrix})P(\begin{smallmatrix} \ast \\ 0 \end{smallmatrix})+pP(\begin{smallmatrix} 1 \\ \ast \end{smallmatrix})P(\begin{smallmatrix} 0 \\ \ast \end{smallmatrix})P(\begin{smallmatrix} \ast \\ 1 \end{smallmatrix})\\
&+\omega_{\rm D1}P(\begin{smallmatrix} 1 \\ \ast \end{smallmatrix})P(\begin{smallmatrix} \ast \\ 0 \end{smallmatrix})+\omega_{\rm D2}P(\begin{smallmatrix} 1 \\ \ast \end{smallmatrix})P(\begin{smallmatrix} \ast \\ 1 \end{smallmatrix})\bigr],
\end{split}
\label{eq:original2}
\end{equation}
where $P(\begin{smallmatrix} 0 \\ \ast \end{smallmatrix})+P(\begin{smallmatrix} 1 \\ \ast \end{smallmatrix})=1$ and $P(\begin{smallmatrix} \ast \\ 0 \end{smallmatrix})+P(\begin{smallmatrix} \ast \\ 1 \end{smallmatrix})=1$. 

Given the obvious symmetry between lanes 1 and 2, after a long enough time we obtain $\rho=P(\begin{smallmatrix} 1 \\ \ast \end{smallmatrix})=P(\begin{smallmatrix} \ast \\ 1 \end{smallmatrix})$; therefore, Eq. (\ref{eq:original2}) reduces to
\begin{equation}
\begin{split}
\frac{d \rho}{dt}=&(\omega_{\rm A1}+\omega_{\rm D1}-\omega_{\rm A2}-\omega_{\rm D2})\rho^2\\
&+(-2\omega_{\rm A1}+\omega_{\rm A2}-\omega_{\rm D1})\rho+\omega_{\rm A1},
\end{split}
\label{eq:original3}
\end{equation}
where all the terms including $p$ disappear.
We note that Eq. (\ref{eq:original3}) does not change for bidirectional flows.

Because $\frac{d\rho}{dt}=0$ in the steady state, we obtain
\begin{equation}
a\rho^2+b\rho+c=0,
\label{eq:original4}
\end{equation}
where
\begin{equation}
a=\omega_{\rm A1}-\omega_{\rm A2}+\omega_{\rm D1}-\omega_{\rm D2},
\end{equation}
\begin{equation}
b=-2\omega_{\rm A1}+\omega_{\rm A2}-\omega_{\rm D1},
\end{equation}
and
\begin{equation}
c=\omega_{\rm A1}.
\end{equation}

For $a\neq0$, the solution of Eq. (\ref{eq:original4}) can be written as 
\begin{equation}
\rho=\frac{-b-\sqrt{b^2-4ac}}{2a},
\label{eq:original5}
\end{equation}
where
\begin{equation}
b^2-4ac=(\omega_{\rm A2}-\omega_{\rm D1})^2+4\omega_{\rm A2}\omega_{\rm D2}>0.
\end{equation}
We discuss the exclusion of the other solution of Eq. (\ref{eq:original4}), i.e., $\rho=\frac{-b+\sqrt{b^2-4ac}}{2a}$, in Appendix \ref{sec:1clusterd}.

However, when $a=0$, we have
\begin{equation}
\rho=\frac{\omega_{\rm A1}}{\omega_{\rm A1}+\omega_{\rm D1}}.
\end{equation}

\subsection{Two-cluster mean-field analysis}
\label{sec:2cluster}

In this subsection, we consider the 2-cluster mean-field analysis, where 2($=2\times1$) sites are regarded as one cluster. We note that this analysis is called “simple mean-field method” in Refs. ~\cite{jiang2009phase,hao2018theoretical,PhysRevE.100.032133}; however, we do not use this terminology in order to clarify the difference from the analysis in the previous subsection and to avoid misunderstanding.

For the 2-cluster probability, the following two master equations can be derived for the case of unidirectional flows:

\begin{equation}
\begin{split}
\frac{d P(\begin{smallmatrix} 0 \\ 0 \end{smallmatrix})}{dt}=&\bigl[P(\begin{smallmatrix}  1  & 0 \\ \underline{0} & 0 \end{smallmatrix})+P(\begin{smallmatrix} 1 & 0 \\  0 & \underline{1}\end{smallmatrix})+P(\begin{smallmatrix}  0  & 0 \\ \underline{1} & 0 \end{smallmatrix})+P(\begin{smallmatrix} 0 & 1 \\  \underline{1} & 0 \end{smallmatrix})\\
&+\omega_{\rm D1}P(\begin{smallmatrix} 1 \\ \underline{0} \end{smallmatrix})+\omega_{\rm D1}P(\begin{smallmatrix} 0 \\ \underline{1} \end{smallmatrix})\bigr]\\
&-\bigl[P(\begin{smallmatrix} 1 & 0 \\  0 & \underline{0}\end{smallmatrix})+P(\begin{smallmatrix} 0 & 0 \\  1 & \underline{0}\end{smallmatrix})+2pP(\begin{smallmatrix} 1 & 0 \\  1 & \underline{0}\end{smallmatrix})\\
&+2\omega_{\rm A1}P(\begin{smallmatrix} 0 \\ \underline{0} \end{smallmatrix})\bigr]
\end{split}
\label{eq:2cluster0}
\end{equation}
and
\begin{equation}
\begin{split}
\frac{d P(\begin{smallmatrix} 1 \\ 0 \end{smallmatrix})}{dt}=&\bigl[P(\begin{smallmatrix} 1 & 0 \\  0 & \underline{0}\end{smallmatrix})+pP(\begin{smallmatrix} 1 & 0 \\  \underline{1} & 0\end{smallmatrix})+pP(\begin{smallmatrix} 1 & 0 \\  1 & \underline{0}\end{smallmatrix})+pP(\begin{smallmatrix} 1 & 1 \\  1 & \underline{0}\end{smallmatrix})\\
&+\omega_{\rm A1}P(\begin{smallmatrix} 0 \\ \underline{0} \end{smallmatrix})+\omega_{\rm D2}P(\begin{smallmatrix} 1 \\ \underline{1} \end{smallmatrix})\bigr]\\
&-\bigl[P(\begin{smallmatrix} 0 & 1 \\  1 & \underline{0}\end{smallmatrix})+pP(\begin{smallmatrix} 1 & 1 \\  1 & \underline{0}\end{smallmatrix})+P(\begin{smallmatrix} 1 & 0 \\  \underline{0} & 0\end{smallmatrix})+P(\begin{smallmatrix} 1 & 0 \\  \underline{0} & 1\end{smallmatrix})\\
&+\omega_{\rm A2}P(\begin{smallmatrix} 1 \\ \underline{0} \end{smallmatrix})+\omega_{\rm D1}P(\begin{smallmatrix} 1 \\ \underline{0} \end{smallmatrix})\bigr].
\end{split}
\label{eq:2cluster1}
\end{equation}


Again performing the mean-field analysis, i.e., ignoring the higher correlations in Eq. (\ref{eq:2cluster0})--(\ref{eq:2cluster1}), we have 
\begin{equation}
\begin{split}
\frac{d P(\begin{smallmatrix} 0 \\ 0 \end{smallmatrix})}{dt}=&\bigl[P(\begin{smallmatrix} 1 \\ 0 \end{smallmatrix})P(\begin{smallmatrix} 0 \\ 0 \end{smallmatrix})+P(\begin{smallmatrix} 1 \\ 0 \end{smallmatrix})P(\begin{smallmatrix} 0 \\ 1 \end{smallmatrix})+P(\begin{smallmatrix} 0 \\ 1 \end{smallmatrix})P(\begin{smallmatrix} 0 \\ 0 \end{smallmatrix})\\
&+P(\begin{smallmatrix} 0 \\ 1 \end{smallmatrix})P(\begin{smallmatrix} 1 \\ 0 \end{smallmatrix})+\omega_{\rm D1}P(\begin{smallmatrix} 1 \\ 0 \end{smallmatrix})+\omega_{\rm D1}P(\begin{smallmatrix} 0 \\ 1 \end{smallmatrix})\bigr]\\
&-\bigl[P(\begin{smallmatrix} 1 \\ 0 \end{smallmatrix})P(\begin{smallmatrix} 0 \\ 0 \end{smallmatrix})+P(\begin{smallmatrix} 0 \\ 1 \end{smallmatrix})P(\begin{smallmatrix} 0 \\ 0 \end{smallmatrix})\\
&+2pP(\begin{smallmatrix} 1 \\ 1 \end{smallmatrix})P(\begin{smallmatrix} 0 \\ 0 \end{smallmatrix})+2\omega_{\rm A1}P(\begin{smallmatrix} 0 \\ 0 \end{smallmatrix})\bigr]\\
=&2P(\begin{smallmatrix} 1 \\ 0 \end{smallmatrix})P(\begin{smallmatrix} 0 \\ 1 \end{smallmatrix})-2pP(\begin{smallmatrix} 0 \\ 0 \end{smallmatrix})P(\begin{smallmatrix} 1 \\ 1 \end{smallmatrix})\\
&+\omega_{\rm D1}P(\begin{smallmatrix} 1 \\ 0 \end{smallmatrix})+\omega_{\rm D1}P(\begin{smallmatrix} 0 \\ 1 \end{smallmatrix})-2\omega_{\rm A1}P(\begin{smallmatrix} 0 \\ 0 \end{smallmatrix})
\end{split}
\label{eq:2cluster00}
\end{equation}
and
\begin{equation}
\begin{split}
\frac{d P(\begin{smallmatrix} 1 \\ 0 \end{smallmatrix})}{dt}=&\bigl[P(\begin{smallmatrix} 1 \\ 0 \end{smallmatrix})P(\begin{smallmatrix} 0 \\ 0 \end{smallmatrix})+pP(\begin{smallmatrix} 1 \\ 1 \end{smallmatrix})P(\begin{smallmatrix} 0 \\ 0 \end{smallmatrix})+pP(\begin{smallmatrix} 1 \\ 1 \end{smallmatrix})P(\begin{smallmatrix} 0 \\ 0 \end{smallmatrix})\\
&+pP(\begin{smallmatrix} 1 \\ 1 \end{smallmatrix})P(\begin{smallmatrix} 1 \\ 0 \end{smallmatrix})+\omega_{\rm A1}P(\begin{smallmatrix} 0 \\ 0 \end{smallmatrix})+\omega_{\rm D2}P(\begin{smallmatrix} 1 \\ 1 \end{smallmatrix})\bigr]\\
&-\bigl[P(\begin{smallmatrix} 0 \\ 1 \end{smallmatrix})P(\begin{smallmatrix} 1 \\ 0 \end{smallmatrix})+pP(\begin{smallmatrix} 1 \\ 1 \end{smallmatrix})P(\begin{smallmatrix} 1 \\ 0 \end{smallmatrix})+P(\begin{smallmatrix} 1 \\ 0 \end{smallmatrix})P(\begin{smallmatrix} 0 \\ 0 \end{smallmatrix})\\
&+P(\begin{smallmatrix} 1 \\ 0 \end{smallmatrix})P(\begin{smallmatrix} 0 \\ 1 \end{smallmatrix})+\omega_{\rm A2}P(\begin{smallmatrix} 1 \\ 0 \end{smallmatrix})+\omega_{\rm D1}P(\begin{smallmatrix} 1 \\ 0 \end{smallmatrix})\bigr]\\
=&2pP(\begin{smallmatrix} 0 \\ 0 \end{smallmatrix})P(\begin{smallmatrix} 1 \\ 1 \end{smallmatrix})-2P(\begin{smallmatrix} 1 \\ 0 \end{smallmatrix})P(\begin{smallmatrix} 0 \\ 1 \end{smallmatrix})+\omega_{\rm A1}P(\begin{smallmatrix} 0 \\ 0 \end{smallmatrix})\\
&+\omega_{\rm D2}P(\begin{smallmatrix} 1 \\ 1 \end{smallmatrix})-\omega_{\rm A2}P(\begin{smallmatrix} 1 \\ 0 \end{smallmatrix})-\omega_{\rm D1}P(\begin{smallmatrix} 1 \\ 0 \end{smallmatrix}).
\end{split}
\label{eq:2cluster11}
\end{equation}

We stress here that Eqs. (\ref{eq:2cluster0}) and (\ref{eq:2cluster1}) change for the case of bidirectional flows; however, Eqs. (\ref{eq:2cluster00}) and (\ref{eq:2cluster11}) do not change and thus yields the same results for the mean-field analysis.

Again given the obvious symmetry between lanes 1 and 2 after a long enough time, we obtain
\begin{equation}
P(\begin{smallmatrix} 1 \\ 0 \end{smallmatrix})=P(\begin{smallmatrix} 0 \\ 1 \end{smallmatrix}).
\end{equation}

Moreover, $P(\begin{smallmatrix} i \\ j \end{smallmatrix})$ ($i,j\in\{0,1\}$) must satisfy the normalization condition:
\begin{equation}
P(\begin{smallmatrix} 0 \\ 0 \end{smallmatrix})+P(\begin{smallmatrix} 1 \\ 0 \end{smallmatrix})+P(\begin{smallmatrix} 0 \\ 1 \end{smallmatrix})+P(\begin{smallmatrix} 1 \\ 1 \end{smallmatrix})=1.
\label{eq:normal}
\end{equation}

Because $\frac{d}{dt}P(\begin{smallmatrix} i \\ j \end{smallmatrix})=0$ in the steady state, we obtain the following expression from Eqs. (\ref{eq:2cluster00})--(\ref{eq:normal}):
\begin{equation}
A\left\{P(\begin{smallmatrix} 1 \\ 0 \end{smallmatrix})\right\}^2+BP(\begin{smallmatrix} 1 \\ 0 \end{smallmatrix})+C=0,
\label{eq:P1eq}
\end{equation}
where
\begin{equation}
A=2+4p\alpha+2p\alpha^2,
\end{equation}
\begin{equation}
B=4p\alpha\beta+4p\beta+\omega_{\rm A2}+\omega_{\rm D1}+2\omega_{\rm D2}+\omega_{\rm D2}\alpha-2p\alpha-\omega_{\rm A1}\alpha,
\end{equation}
\begin{equation}
C=2p\beta^2+\omega_{\rm D2}\beta-2p\beta-\omega_{\rm A1}\beta-\omega_{\rm D2},
\end{equation}
\begin{equation}
\alpha=\frac{\omega_{\rm D1}-\omega_{\rm A2}-2\omega_{\rm D2}}{\omega_{\rm A1}+\omega_{\rm D2}},
\label{eq:alpha}
\end{equation}
and
\begin{equation}
\beta=\frac{\omega_{\rm D2}}{\omega_{\rm A1}+\omega_{\rm D2}}.
\label{eq:beta}
\end{equation}
Solving Eq. (\ref{eq:P1eq}) yields $P(\begin{smallmatrix} 1 \\ 0 \end{smallmatrix})$ in the form 
\begin{equation}
P(\begin{smallmatrix} 1 \\ 0 \end{smallmatrix})=\frac{-B+\sqrt{B^2-4AC}}{2A}.
\label{eq:P1}
\end{equation}
We note that because
\begin{equation}
A=2+4p\alpha+2p\alpha^2=2(1-p)+2p(\alpha+1)^2>0
\label{eq:A0}
\end{equation}
and
\begin{eqnarray}
C&=&2p\beta^2+\omega_{\rm D2}\beta-2p\beta-\omega_{\rm A1}\beta-\omega_{\rm D2}\\
&=&-\frac{2p\omega_{\rm A1}-\omega_{\rm D2}\omega_{\rm A1}-\omega_{\rm A1}\omega_{\rm D2}}{\omega_{\rm A1}+\omega_{\rm D2}}<0,
\label{eq:C0}
\end{eqnarray}
we have
\begin{equation}
B^2-4AC>0.
\end{equation}
We discuss the exclusion of the other solution of Eq. (\ref{eq:P1eq}), i.e., $P(\begin{smallmatrix} 1 \\ 0 \end{smallmatrix})=\frac{-B-\sqrt{B^2-4AC}}{2A}$ in Appendix \ref{sec:2clusterd}.

Because the density is defined as
\begin{equation}
\rho=P(\begin{smallmatrix} 1 \\ 0 \end{smallmatrix})+P(\begin{smallmatrix} 1 \\ 1 \end{smallmatrix}),
\end{equation}
we finally have
\begin{equation}
\rho=1-\beta-\frac{(1+\alpha)(-B+\sqrt{B^2-4AC})}{2A}.
\label{eq:rho2cluster}
\end{equation}

\subsection{Four-cluster mean-field analysis}
\label{sec:4cluster}

This subsection presents the 4-cluster mean-field analysis, where 4($=2\times2$) sites are regarded as one cluster. We note that this analysis is called the “2-cluster mean-field method” in Refs. ~\cite{jiang2009phase,hao2018theoretical,PhysRevE.100.032133}; however, as with Subsec. \ref{sec:2cluster}, we do not use this terminology in order to clarify the difference from the analyses in the last two subsections and to avoid misunderstanding.

Unlike the two previous mean-field analyses, in this case the final results are different for the two directional flows. Therefore, in this subsection we consider the two flows separately.

\subsubsection{Unidirectional flows}

The master equation for $P(\begin{smallmatrix} 0 & 0 \\  0 & 0 \end{smallmatrix})$ can be expressed as

\begin{equation}
\begin{split}
\frac{d P(\begin{smallmatrix} 0 & 0 \\  0 & 0 \end{smallmatrix})}{dt}=&\bigl[P(\begin{smallmatrix} 0 & 1 & 0 \\  \underline{0} & \underline{0} &0  \end{smallmatrix})+P(\begin{smallmatrix} 0 & 1 & 0 \\  \underline{0} & \underline{0} & 1  \end{smallmatrix})+P(\begin{smallmatrix} 0 & 0 & 0 \\  \underline{0} & \underline{1} & 0  \end{smallmatrix})\\
&+P(\begin{smallmatrix} 0 & 0 & 1 \\  \underline{0} & \underline{1} & 0  \end{smallmatrix})+\omega_{\rm D1}P(\begin{smallmatrix} 1 & 0 \\  \underline{0} & \underline{0} \end{smallmatrix})+\omega_{\rm D1}P(\begin{smallmatrix} 0 & 0 \\  \underline{1} & \underline{0} \end{smallmatrix})\\
&+\omega_{\rm D1}P(\begin{smallmatrix} 0 & 1 \\  \underline{0} & \underline{0} \end{smallmatrix})+\omega_{\rm D1}P(\begin{smallmatrix} 0 & 0 \\  \underline{0} & \underline{1} \end{smallmatrix})\bigr]\\
&-\bigl[P(\begin{smallmatrix} 1 & 0 & 0 \\  0 & \underline{0} & \underline{0}  \end{smallmatrix})+P(\begin{smallmatrix} 0 & 0 & 0 \\  1 & \underline{0} & \underline{0}  \end{smallmatrix})+2pP(\begin{smallmatrix} 1 & 0 & 0 \\  1 & \underline{0} & \underline{0}  \end{smallmatrix})\\
&+4\omega_{\rm A1}P(\begin{smallmatrix} 0 & 0 \\  \underline{0} & \underline{0} \end{smallmatrix})\bigr].\\
\end{split}
\label{eq:4cluster00}
\end{equation}

Utilizing the concept of conditional probability, we can express $P(\begin{smallmatrix} i & k & m \\  j & l & n  \end{smallmatrix})$ in this mean-field analysis in the form
\begin{equation}
P(\begin{smallmatrix} i & k & m \\  j & l & n  \end{smallmatrix})=\frac{P(\begin{smallmatrix} i & k \\ j & l \end{smallmatrix})P(\begin{smallmatrix} k & m \\ l & n \end{smallmatrix})}{\sum_{m\in\{0,1\}}\sum_{n\in\{0,1\}} P(\begin{smallmatrix} k & m \\ l & n \end{smallmatrix})},
\label{eq:condition}
\end{equation}
where $i,j,k,l,m,n\in\{0,1\}$.

Inserting Eq. (\ref{eq:condition}) into Eq. (\ref{eq:4cluster00}) and noting that in the steady state $\frac{d}{dt}P(\begin{smallmatrix} i & k \\  j & l \end{smallmatrix})=0$, we obtain

\begin{equation}
\begin{split}
&\frac{P(\begin{smallmatrix} 0 & 1 \\  0 & 0  \end{smallmatrix})P(\begin{smallmatrix} 1 & 0 \\  0 & 0  \end{smallmatrix})}{\sum_{i\in\{0,1\}}\sum_{j\in\{0,1\}} P(\begin{smallmatrix} 1 & i \\ 0 & j \end{smallmatrix})}+\frac{P(\begin{smallmatrix} 0 & 1 \\  0 & 0  \end{smallmatrix})P(\begin{smallmatrix} 1 & 0 \\  0 & 1  \end{smallmatrix})}{\sum_{i\in\{0,1\}}\sum_{j\in\{0,1\}} P(\begin{smallmatrix} 1 & i \\ 0 & j \end{smallmatrix})}\\
&+\frac{P(\begin{smallmatrix} 0 & 0 \\  0 & 1  \end{smallmatrix})P(\begin{smallmatrix} 0 & 0 \\  1 & 0  \end{smallmatrix})}{\sum_{i\in\{0,1\}}\sum_{j\in\{0,1\}} P(\begin{smallmatrix} 0 & i \\ 1 & j \end{smallmatrix})}+\frac{P(\begin{smallmatrix} 0 & 0 \\  0 & 1  \end{smallmatrix})P(\begin{smallmatrix} 0 & 1 \\  1 & 0  \end{smallmatrix})}{\sum_{i\in\{0,1\}}\sum_{j\in\{0,1\}} P(\begin{smallmatrix} 0 & i \\ 1 & j \end{smallmatrix})}\\
&+\omega_{\rm D1}P(\begin{smallmatrix} 1 & 0 \\  0 & 0 \end{smallmatrix})+\omega_{\rm D1}P(\begin{smallmatrix} 0 & 0 \\  1 & 0 \end{smallmatrix})+\omega_{\rm D1}P(\begin{smallmatrix} 0 & 1 \\  0 & 0 \end{smallmatrix})+\omega_{\rm D1}P(\begin{smallmatrix} 0 & 0 \\  0 & 1 \end{smallmatrix})\\
&-\frac{P(\begin{smallmatrix} 1 & 0 \\  0 & 0  \end{smallmatrix})P(\begin{smallmatrix} 0 & 0 \\  0 & 0  \end{smallmatrix})}{\sum_{i\in\{0,1\}}\sum_{j\in\{0,1\}} P(\begin{smallmatrix} 0 & i \\ 0 & j \end{smallmatrix})}-\frac{P(\begin{smallmatrix} 0 & 0 \\  1 & 0  \end{smallmatrix})P(\begin{smallmatrix} 0 & 0 \\  0 & 0  \end{smallmatrix})}{\sum_{i\in\{0,1\}}\sum_{j\in\{0,1\}} P(\begin{smallmatrix} 0 & i \\ 0 & j \end{smallmatrix})}\\
&-\frac{2pP(\begin{smallmatrix} 1 & 0 \\  1 & 0  \end{smallmatrix})P(\begin{smallmatrix} 0 & 0 \\  0 & 0  \end{smallmatrix})}{\sum_{i\in\{0,1\}}\sum_{j\in\{0,1\}} P(\begin{smallmatrix} 0 & i \\ 0 & j \end{smallmatrix})}-4\omega_{\rm A1}P(\begin{smallmatrix} 0 & 0 \\  0 & 0 \end{smallmatrix})=0.
\end{split}
\label{eq:4cluster00s}
\end{equation}

We can obtain the eight other master equations for the 4-cluster probabilities in the steady state similarly, as shown in Appendix \ref{sec:4othermasteruni}.

In addition, given the obvious symmetry between lanes 1 and 2 after a long enough time, we obtain
\begin{equation}
P(\begin{smallmatrix} 1 & 0 \\ 0 & 0 \end{smallmatrix})=P(\begin{smallmatrix} 0 & 0 \\ 1 & 0 \end{smallmatrix}),
\end{equation}
\begin{equation}
P(\begin{smallmatrix} 0 & 1 \\ 0 & 0 \end{smallmatrix})=P(\begin{smallmatrix} 0 & 0 \\ 0 & 1 \end{smallmatrix}),
\end{equation}
\begin{equation}
P(\begin{smallmatrix} 1 & 1 \\ 0 & 0 \end{smallmatrix})=P(\begin{smallmatrix} 0 & 0 \\ 1 & 1 \end{smallmatrix}),
\end{equation}
\begin{equation}
P(\begin{smallmatrix} 1 & 0 \\ 0 & 1 \end{smallmatrix})=P(\begin{smallmatrix} 0 & 1 \\ 1 & 0 \end{smallmatrix}),
\end{equation}
\begin{equation}
P(\begin{smallmatrix} 1 & 1 \\ 1 & 0 \end{smallmatrix})=P(\begin{smallmatrix} 1 & 0 \\ 1 & 1 \end{smallmatrix}),
\end{equation}
and
\begin{equation}
P(\begin{smallmatrix} 1 & 1 \\ 0 & 1 \end{smallmatrix})=P(\begin{smallmatrix} 0 & 1 \\ 1 & 1 \end{smallmatrix}).
\end{equation}

Moreover, $P(\begin{smallmatrix} i & k \\  j & l \end{smallmatrix})$ must satisfy the normalization condition
\begin{equation}
\sum_{i\in\{0,1\}}\sum_{j\in\{0,1\}}\sum_{k\in\{0,1\}}\sum_{l\in\{0,1\}}P(\begin{smallmatrix} i & k \\  j & l \end{smallmatrix})=1.
\label{eq:normal4}
\end{equation}

From 16 independent Eqs. (\ref{eq:4cluster00s})--(\ref{eq:normal4}) and (\ref{eq:4cluster302})--(\ref{eq:4cluster132}), we obtain all the probabilities $P(\begin{smallmatrix} i & k \\  j & l \end{smallmatrix})$. 

Finally, the definition of density gives
\begin{equation}
\rho=\sum_{i\in\{0,1\}}\sum_{j\in\{0,1\}}\sum_{k\in\{0,1\}}P(\begin{smallmatrix} 1 & j \\  i & k \end{smallmatrix}).
\label{eq:rho4cluster}
\end{equation}

\subsubsection{Bidirectional flows}

The master equation for $P(\begin{smallmatrix} 0 & 0 \\  0 & 0 \end{smallmatrix})$ can be written in the form

\begin{equation}
\begin{split}
\frac{d P(\begin{smallmatrix} 0 & 0 \\  0 & 0 \end{smallmatrix})}{dt}=&\bigl[P(\begin{smallmatrix} 0 & 0 & 0 \\  0 & \underline{1} & \underline{0}  \end{smallmatrix})+P(\begin{smallmatrix} 1 & 0 & 0 \\  0 & \underline{1} & \underline{0}  \end{smallmatrix})+P(\begin{smallmatrix} 0 & 1 & 0 \\  \underline{0} & \underline{0} & 0  \end{smallmatrix})\\
&+P(\begin{smallmatrix} 0 & 1 & 0 \\  \underline{0} & \underline{0} & 1  \end{smallmatrix})+\omega_{\rm D1}P(\begin{smallmatrix} 1 & 0 \\  \underline{0} & \underline{0} \end{smallmatrix})+\omega_{\rm D1}P(\begin{smallmatrix} 0 & 0 \\  \underline{1} & \underline{0} \end{smallmatrix})\\
&+\omega_{\rm D1}P(\begin{smallmatrix} 0 & 1 \\  \underline{0} & \underline{0} \end{smallmatrix})+\omega_{\rm D1}P(\begin{smallmatrix} 0 & 0 \\  \underline{0} & \underline{1} \end{smallmatrix})\bigr]\\
&-\bigl[P(\begin{smallmatrix} 1 & 0 & 0 \\  0 & \underline{0} & \underline{0}  \end{smallmatrix})+pP(\begin{smallmatrix} 1 & 0 & 0 \\  1 & \underline{0} & \underline{0}  \end{smallmatrix})+P(\begin{smallmatrix} 0 & 0 & 0 \\  \underline{0} & \underline{0} & 1  \end{smallmatrix})\\
&+pP(\begin{smallmatrix} 0 & 0 & 1 \\  \underline{0} & \underline{0} & 1  \end{smallmatrix})+4\omega_{\rm A1}P(\begin{smallmatrix} 0 & 0 \\  \underline{0} & \underline{0} \end{smallmatrix})\bigr].\\
\end{split}
\label{eq:4cluster00b}
\end{equation}

Inserting Eq. (\ref{eq:condition}) into Eq. (\ref{eq:4cluster00b}) and noting that in the steady state $\frac{d}{dt}P(\begin{smallmatrix} i & k \\  j & l \end{smallmatrix})=0$, we obtain

\begin{equation}
\begin{split}
&\frac{P(\begin{smallmatrix} 0 & 0 \\  0 & 1  \end{smallmatrix})P(\begin{smallmatrix} 0 & 0 \\  1 & 0  \end{smallmatrix})}{\sum_{i\in\{0,1\}}\sum_{j\in\{0,1\}} P(\begin{smallmatrix} 0 & i \\ 1 & j \end{smallmatrix})}+\frac{P(\begin{smallmatrix} 1 & 0 \\  0 & 1  \end{smallmatrix})P(\begin{smallmatrix} 0 & 0 \\  1 & 0  \end{smallmatrix})}{\sum_{i\in\{0,1\}}\sum_{j\in\{0,1\}} P(\begin{smallmatrix} 0 & i \\ 1 & j \end{smallmatrix})}\\
&+\frac{P(\begin{smallmatrix} 0 & 1 \\  0 & 0  \end{smallmatrix})P(\begin{smallmatrix} 1 & 0 \\  0 & 0  \end{smallmatrix})}{\sum_{i\in\{0,1\}}\sum_{j\in\{0,1\}} P(\begin{smallmatrix} 1 & i \\ 0 & j \end{smallmatrix})}+\frac{P(\begin{smallmatrix} 0 & 1 \\  0 & 0  \end{smallmatrix})P(\begin{smallmatrix} 1 & 0 \\  0 & 1  \end{smallmatrix})}{\sum_{i\in\{0,1\}}\sum_{j\in\{0,1\}} P(\begin{smallmatrix} 1 & i \\ 0 & j \end{smallmatrix})}\\
&+\omega_{\rm D1}P(\begin{smallmatrix} 1 & 0 \\  0 & 0 \end{smallmatrix})+\omega_{\rm D1}P(\begin{smallmatrix} 0 & 0 \\  1 & 0 \end{smallmatrix})+\omega_{\rm D1}P(\begin{smallmatrix} 0 & 1 \\  0 & 0 \end{smallmatrix})+\omega_{\rm D1}P(\begin{smallmatrix} 0 & 0 \\  0 & 1 \end{smallmatrix})\\
&-\frac{P(\begin{smallmatrix} 1 & 0 \\  0 & 0  \end{smallmatrix})P(\begin{smallmatrix} 0 & 0 \\  0 & 0  \end{smallmatrix})}{\sum_{i\in\{0,1\}}\sum_{j\in\{0,1\}} P(\begin{smallmatrix} 0 & i \\ 0 & j \end{smallmatrix})}-\frac{pP(\begin{smallmatrix} 1 & 0 \\  1 & 0  \end{smallmatrix})P(\begin{smallmatrix} 0 & 0 \\  0 & 0  \end{smallmatrix})}{\sum_{i\in\{0,1\}}\sum_{j\in\{0,1\}} P(\begin{smallmatrix} 0 & i \\ 0 & j \end{smallmatrix})}\\
&-\frac{P(\begin{smallmatrix} 0 & 0 \\  0 & 0  \end{smallmatrix})P(\begin{smallmatrix} 0 & 0 \\  0 & 1  \end{smallmatrix})}{\sum_{i\in\{0,1\}}\sum_{j\in\{0,1\}} P(\begin{smallmatrix} 0 & i \\ 0 & j \end{smallmatrix})}-\frac{pP(\begin{smallmatrix} 0 & 0 \\  0 & 0  \end{smallmatrix})P(\begin{smallmatrix} 0 & 1 \\  0 & 1  \end{smallmatrix})}{\sum_{i\in\{0,1\}}\sum_{j\in\{0,1\}} P(\begin{smallmatrix} 0 & i \\ 0 & j \end{smallmatrix})}\\
&-4\omega_{\rm A1}P(\begin{smallmatrix} 0 & 0 \\  0 & 0 \end{smallmatrix})=0.
\end{split}
\label{eq:4cluster00bd}
\end{equation}

We can again obtain the eight other master equations for the 4-cluster probabilities in the steady state similarly, as shown in Appendix \ref{sec:4othermasterbi}.

Given the obvious symmetry between lanes 1 and 2 after a long enough time, we obtain
\begin{equation}
P(\begin{smallmatrix} 1 & 0 \\ 0 & 0 \end{smallmatrix})=P(\begin{smallmatrix} 0 & 0 \\ 0 & 1 \end{smallmatrix}),
\end{equation}
\begin{equation}
P(\begin{smallmatrix} 0 & 0 \\ 1 & 0 \end{smallmatrix})=P(\begin{smallmatrix} 0 & 1 \\ 0 & 0 \end{smallmatrix}),
\end{equation}
\begin{equation}
P(\begin{smallmatrix} 1 & 1 \\ 0 & 0 \end{smallmatrix})=P(\begin{smallmatrix} 0 & 0 \\ 1 & 1 \end{smallmatrix}),
\end{equation}
\begin{equation}
P(\begin{smallmatrix} 1 & 0 \\ 1 & 0 \end{smallmatrix})=P(\begin{smallmatrix} 0 & 1 \\ 0 & 1 \end{smallmatrix}),
\end{equation}
\begin{equation}
P(\begin{smallmatrix} 1 & 1 \\ 1 & 0 \end{smallmatrix})=P(\begin{smallmatrix} 0 & 1 \\ 1 & 1 \end{smallmatrix}),
\end{equation}
and
\begin{equation}
P(\begin{smallmatrix} 1 & 0 \\ 1 & 1 \end{smallmatrix})=P(\begin{smallmatrix} 1 & 1 \\ 0 & 1 \end{smallmatrix}).
\label{eq:symmetry}
\end{equation}

From 16 independent Eqs. (\ref{eq:normal4}), (\ref{eq:4cluster00bd})--(\ref{eq:symmetry}), and (\ref{eq:4cluster122b})--(\ref{eq:4cluster322b}), we obtain all the probabilities $P(\begin{smallmatrix} i & k \\  j & l \end{smallmatrix})$. 

Finally, the definition of density gives Eq. (\ref{eq:rho4cluster}).

\section{COMPARISON OF NUMERICAL RESULTS WITH MEAN-FIELD ANALYSES AND SIMULATION RESULTS}
\label{sec:comparison}

In this section, we compare numerical results from the three (1-, 2-, and 4-cluster) mean-field analyses with simulation results. Although analytical solutions can be derived for 1- and 2-cluster mean-field analyses, similar solutions  cannot be obtained explicitly for the 4-cluster mean-field analysis (see the last section). We therefore obtain numerical solutions for this case using Newton's iteration method. In this study, we use the function FindRoot, which is based on the Newton's method, in the software package Mathematica 12.0. In all the simulations below, we set $L=100$ and calculate the steady-state value of $\rho$ for $10^7$ time steps after evolving the system for $10^7$ time steps, unless otherwise specified. 

\subsection{Special case: $\omega_{\rm A1}=\omega_{\rm A2}$ and $\omega_{\rm D1}=\omega_{\rm D2}$}
\label{sec:special}

In this special case, the results of all three mean-field analyses for the two directional flows give us the following equation for the steady-state value of $\rho$:
\begin{equation}
\rho=\frac{\omega_{\rm A}}{\omega_{\rm A}+\omega_{\rm D}}.
\label{eq:rhospecial}
\end{equation}
Eq. (\ref{eq:rhospecial}) shows that all the mean-field analyses give a result independent of $p$ for this special cases. 

\begin{figure*}[htbp]
\begin{center}
\includegraphics[width=16.5cm,clip]{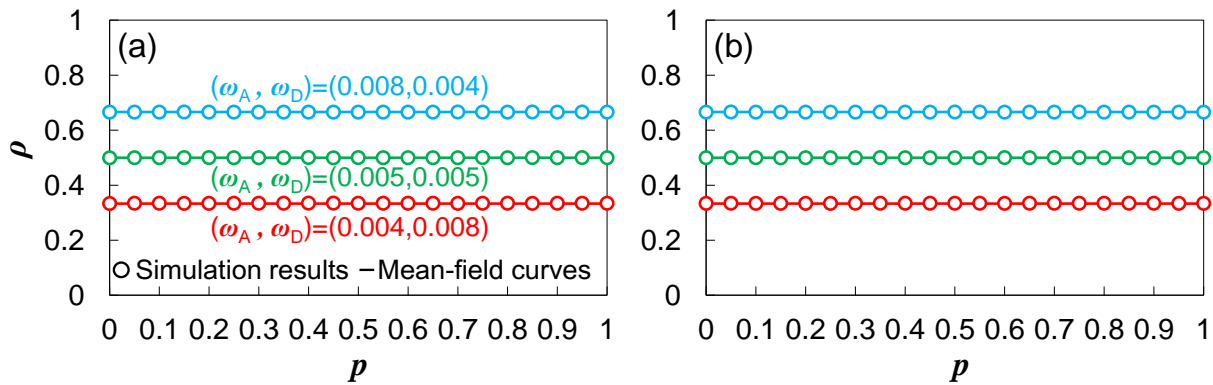}
\caption{(Color Online) Simulation (circles) and mean-field values (curves) of $\rho$ for (a) unidirectional and (b) bidirectional flows as functions of $p$ with $(\omega_{\rm A},\omega_{\rm D})\in\{(0.004,0.008) \ ({\rm red}), (0.005,0.005) \ ({\rm green}), (0.008,0.004) \ ({\rm blue})\}$.}
\label{fig:special}
\end{center}
\end{figure*}

Figure \ref{fig:special} compares the simulation and mean-field values of $\rho$ as functions of $p$ for various $(\omega_{\rm A},\omega_{\rm D})\in\{(0.004,0.008),(0.005,0.005),(0.008,0.004)\}$ for (a) unidirectional and (b) bidirectional flows. In both figures, the simulations show very good agreement with our mean-field analyses.

\subsection{General case}
In this subsection, we consider the more general case with $\omega_{\rm A1}\neq\omega_{\rm A2}$ or $\omega_{\rm D1}\neq\omega_{\rm D2}$. 

Table \ref{tab:comparison} summarizes the three kinds of mean-field analyses (1-, 2-, and 4-cluster cases), which were discussed in detail in Sec. \ref{sec:mean}. In the 1-cluster mean-field analysis, $\rho$ is a function of $\omega$ independent of $p$; therefore, the influence of $p$ cannot be captured. In contrast, the 2-cluster mean-field analysis gives $\rho$ as a function of both $\omega$ and $p$; however, the function is the same for unidirectional and bidirectional flows. Finally, in the 4-cluster mean-field analysis, $\rho$ is a function of both $\omega$ and $p$, but the function is different for unidirectional and bidirectional flows. From this discussion, we expect that the 1-cluster mean-field analysis to approximate the simulation results well in cases where $p\sim1$, whereas the 2-cluster mean-field analysis roughly captures the influence of $p$, and the 4-cluster mean-field analysis reproduces the difference between two directional flows.

\begin{table}[htbp]
\centering
\caption{Comparison of the three kinds of mean-field analyses.}
\label{tab:comparison}
\begin{tabular}{c|c|c}
Mean-field analysis & Direction & $p$ \\ \hline \hline
1-cluster & Independent & Independent \\ \hline
2-cluster & Independent & Dependent\\ \hline
4-cluster & Dependent & Dependent \\
\end{tabular}
\end{table}

\begin{figure*}[htbp]
\begin{center}
\includegraphics[width=16cm,clip]{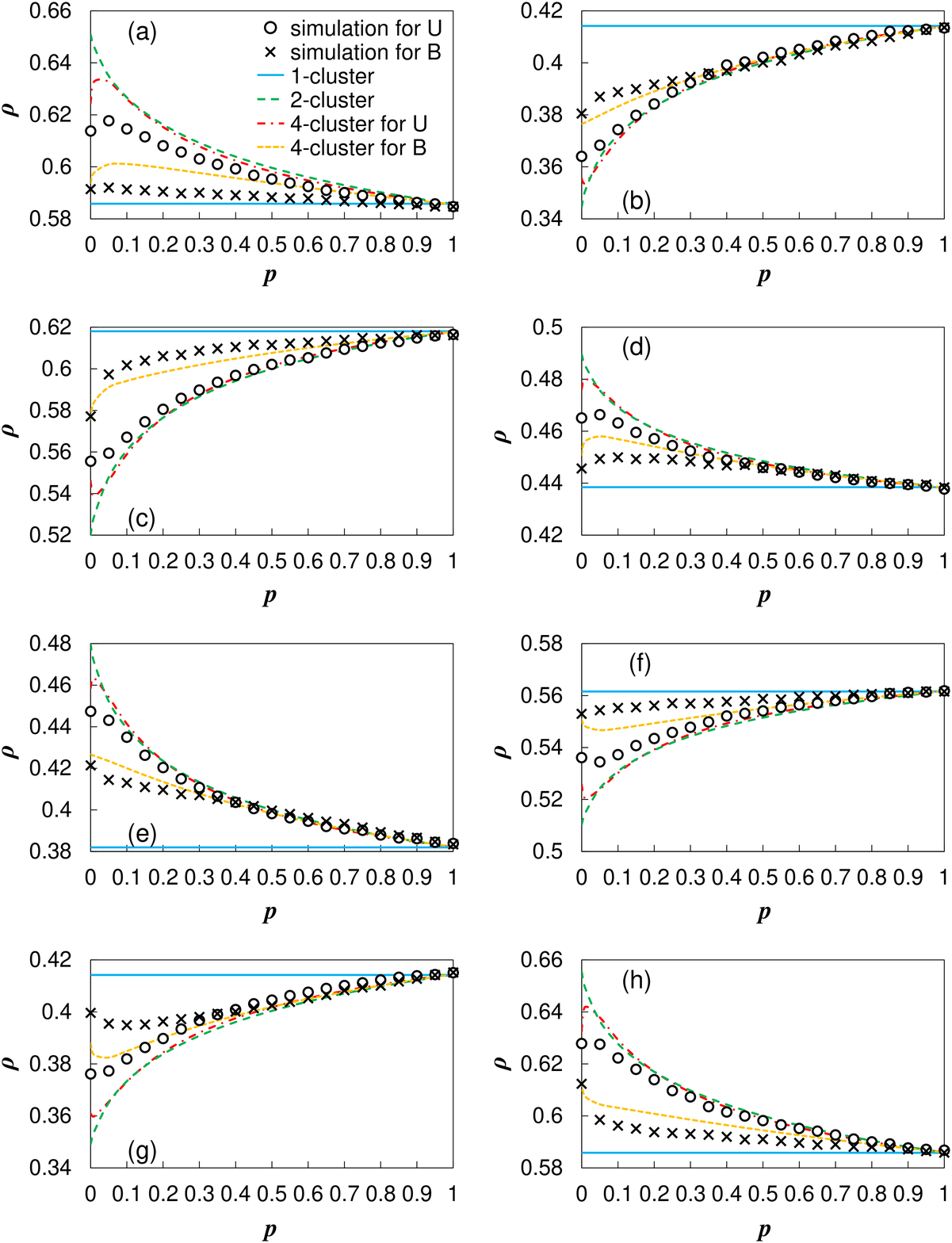}
\caption{(Color Online) Values of $\rho$ from simulations (symbols) and mean-field calculations (blue solid/green dashed/red dash-dotted/orange dotted curves) as functions of $p$ for (a) $(\omega_{\rm A1},\omega_{\rm A2},\omega_{\rm D1},\omega_{\rm D2})=(0.008,0.004,0.004,0.004)$, (b) $(\omega_{\rm A1},\omega_{\rm A2},\omega_{\rm D1},\omega_{\rm D2})=(0.002,0.004,0.004,0.004)$, (c) $(\omega_{\rm A1},\omega_{\rm A2},\omega_{\rm D1},\omega_{\rm D2})=(0.004,0.008,0.004,0.004)$, (d) $(\omega_{\rm A1},\omega_{\rm A2},\omega_{\rm D1},\omega_{\rm D2})=(0.004,0.002,0.004,0.004)$, (e) $(\omega_{\rm A1},\omega_{\rm A2},\omega_{\rm D1},\omega_{\rm D2})=(0.004,0.004,0.008,0.004)$, (f) $(\omega_{\rm A1},\omega_{\rm A2},\omega_{\rm D1},\omega_{\rm D2})=(0.004,0.004,0.002,0.004)$, (g) $(\omega_{\rm A1},\omega_{\rm A2},\omega_{\rm D1},\omega_{\rm D2})=(0.004,0.004,0.004,0.008)$, and (h) $(\omega_{\rm A1},\omega_{\rm A2},\omega_{\rm D1},\omega_{\rm D2})=(0.004,0.004,0.004,0.002)$. Black circles represent the simulation results for unidirectional flows (U), whereas black crosses represent those for bidirectional flows (B). The blue/green/red/orange curves represent the 1-cluster/2-cluster/4-cluster (U)/4-cluster (B) mean-field values, respectively.}
\label{fig:general}
\end{center}
\end{figure*}

We next consider eight fundamental cases; specifically,
\begin{equation}
  \left\{
\begin{array}{ll}
 {\rm (a)} \ \ (\omega_{\rm A1},\omega_{\rm A2},\omega_{\rm D1},\omega_{\rm D2})=(0.008,0.004,0.004,0.004). \\
 {\rm (b)} \ \ (\omega_{\rm A1},\omega_{\rm A2},\omega_{\rm D1},\omega_{\rm D2})=(0.002,0.004,0.004,0.004). \\
 {\rm (c)} \ \ (\omega_{\rm A1},\omega_{\rm A2},\omega_{\rm D1},\omega_{\rm D2})=(0.004,0.008,0.004,0.004). \\
 {\rm (d)} \ \ (\omega_{\rm A1},\omega_{\rm A2},\omega_{\rm D1},\omega_{\rm D2})=(0.004,0.002,0.004,0.004).\\
 {\rm (e)} \ \ (\omega_{\rm A1},\omega_{\rm A2},\omega_{\rm D1},\omega_{\rm D2})=(0.004,0.004,0.008,0.004). \\
 {\rm (f)} \ \ (\omega_{\rm A1},\omega_{\rm A2},\omega_{\rm D1},\omega_{\rm D2})=(0.004,0.004,0.002,0.004).\\
 {\rm (g)} \ \ (\omega_{\rm A1},\omega_{\rm A2},\omega_{\rm D1},\omega_{\rm D2})=(0.004,0.004,0.004,0.008). \\
 {\rm (h)} \ \ (\omega_{\rm A1},\omega_{\rm A2},\omega_{\rm D1},\omega_{\rm D2})=(0.004,0.004,0.004,0.002).
\end{array}
\right.
\end{equation}
These cases enable us to investigate the influence of changes in $\omega_{\rm A}$ or $\omega_{\rm D}$ depending on the state of the corresponding site in the other lane. Specifically, Cases (a)--(d) exhibit the influence of $\omega_{\rm A}$, whereas Cases (e)--(h) show that of $\omega_{\rm D}$. Figure \ref{fig:general} plots the simulation and mean-field curves of $\rho$ for the parameter sets of Cases (a)--(h). 

\begin{figure*}[htbp]
\begin{center}
\includegraphics[width=17.5cm,clip]{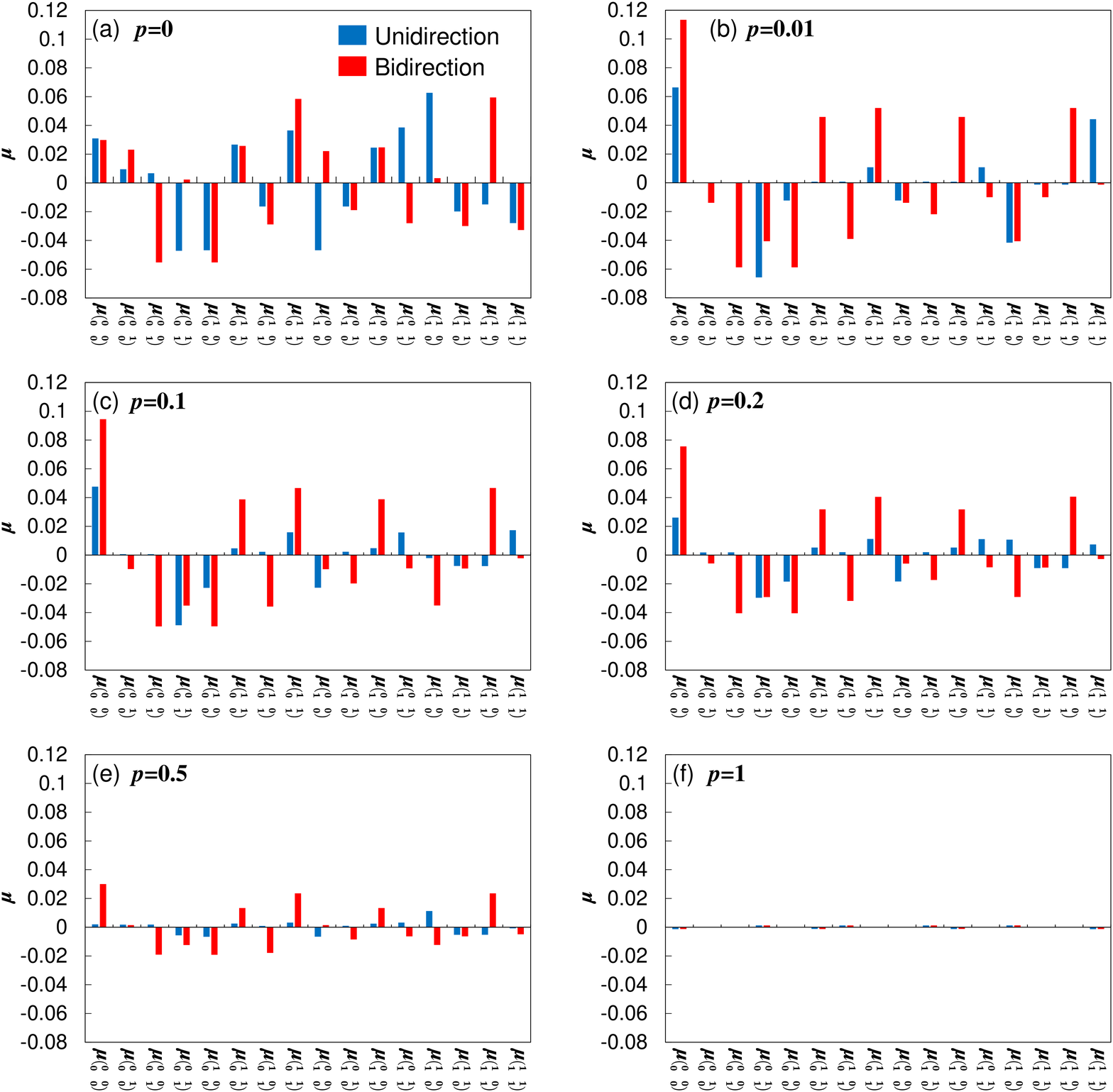}
\caption{(Color Online) Calculated values of 16 kinds of $\mu$ of unidirectional (blue) and bidirectional (red) flows for various $p\in\{0,0.01,0.1,0.2,0.5,1\}$, fixing $(\omega_{\rm A1},\omega_{\rm A2},\omega_{\rm D1},\omega_{\rm D2})=(0,0,0,0)$ and $\rho=0.5$. We note (i) that those values are calculated by averaging 100 different initial configurations for $p=0$ and (ii) that they are calculated for $10^9$ time steps after evolving the system with $p=0.01$ for $10^9$ time steps because it takes more time for the system to evolve into the steady state.}
\label{fig:mu}
\end{center}
\end{figure*}

As expected, Fig. \ref{fig:general}(a)--(h) shows that (i) the 1-cluster mean-field analysis agrees very well with the simulation results only near $p=1$, (ii) the 2-cluster mean-field results capture well the qualitative change of $\rho$ with $(p,\omega)$ for both directions, and (iii) the 4-cluster case not only improves the accuracy of the approximations but also succeeds in reproducing the difference between the two directional flows. We note that the deviation of the numerical results from simulations is the smallest for 1-cluster mean-field analysis, depending on $\omega$, e.g., Cases (a) and (f), although the 1-cluster analysis does not capture the $p$-dependence.

In addition, we observe an interesting phenomenon in Fig. \ref{fig:general}(a)--(h). Specifically, there are smaller discrepancies between the simulations and numerical results for the 2- and 4-cluster mean-field analyses for unidirectional flows than for bidirectional flows. In other words, the numerical results from the 2-cluster mean-field analysis are already good approximations for unidirectional flows.

In the following subsections, we discuss in detail the discrepancies between the numerical results for the 2- and 4-cluster mean-field analyses and for $(p,\omega)$-dependence of $\rho$ for the two directional flows.

\subsubsection{Discrepancies between the numerical results from the 2- and 4-cluster mean-field analyses}

Figure \ref{fig:general} shows that the discrepancies are smaller for unidirectional flows than for bidirectional flows.

To investigate this phenomenon, we first define the following correlation between two adjacent clusters, each of which consists of $2(=2\times1)$ sites:
\begin{equation}
\mu(\begin{smallmatrix} i & k \\ j & l \end{smallmatrix})=P\left(\begin{smallmatrix} i & k \\ j & l \end{smallmatrix}\right)-P\left(\begin{smallmatrix} i \\ j\end{smallmatrix}\right)P\left(\begin{smallmatrix} k \\ l\end{smallmatrix}\right),
\end{equation}
where $i,j,k,l\in\{0,1\}$. From the definition, $|\mu|=0$ indicates that there is no correlation between the two adjacent clusters, whereas $\mu>0$ ($\mu<0$) indicates that the possibility of the spontaneous appearance of two adjacent clusters is larger (smaller) than the possibility under the assumption that clusters appear randomly in the system. 
This explains why the 4-cluster mean-field analysis improves the approximate accuracy more than the 2-cluster analysis for relatively large $|\mu|$. In contrast, the 2-cluster analysis is already a good approximation for relatively small $|\mu|$.

Figure \ref{fig:mu} compares 16 kinds of $\mu$ for unidirectional and bidirectional flows for various $p\in\{0,0.01,0.1,0.2,0.5,1\}$, fixing $(\omega_{\rm A1},\omega_{\rm A2},\omega_{\rm D1},\omega_{\rm D2})=(0,0,0,0)$ and $\rho=0.5$. We set $(\omega_{\rm A1},\omega_{\rm A2},\omega_{\rm D1},\omega_{\rm D2})=(0,0,0,0)$ to observe the pure correlations caused by $p$ because Langmuir kinetics reduce the correlations. We note that---strictly speaking---we cannot discuss the case $p=0$ in the same way as for the cases with $p>0$ because in the former case, all the particles stop at some point, depending on the initial configurations, due to the blocking effect (discussed later). Therefore, Fig. \ref{fig:mu}(a) is just presented for  reference (as are the points at $p=0$ in Fig. \ref{fig:P1P2P3}).  

We observe two important phenomena in Fig. \ref{fig:mu}.
First, for the two directional flows $|\mu|$ approaches to 0 as $p$ increases; i.e., the correlation becomes smaller with larger $p$. This indicates that for larger $p$, the discrepancies between the numerical results from the 2- and 4-cluster mean-field analyses become smaller, which explains the overlap of the 2- and 4-cluster theoretical lines in Fig. \ref{fig:general}.

Second, we confirm that in most cases many values of $|\mu|$ for unidirectional flows are smaller than those for bidirectional flows with the same values of $p$ [Fig. \ref{fig:mu}(b)--(e)]; i.e., the correlations for unidirectional flows are smaller than those for bidirectional flows. This indicates that there are smaller discrepancies between the numerical results from the 2- and 4-cluster mean-field analyses for unidirectional flows compared with those for bidirectional flows. In contrast, for $p=0$ [Fig. \ref{fig:mu}(a)], some values of $|\mu|$ become large even for unidirectional flows, and therefore, the discrepancies also become large (see Fig. \ref{fig:general}).

\subsubsection{$(p,\omega)$-dependence of $\rho$}

This subsection discusses the $(p,\omega)$-dependence of $\rho$, observed from Fig. \ref{fig:general}, for the numerical results from the 2- and 4-cluster mean-field analyses and for the simulation results.

\begin{figure}[htbp]
\begin{center}
\includegraphics[width=8.5cm,clip]{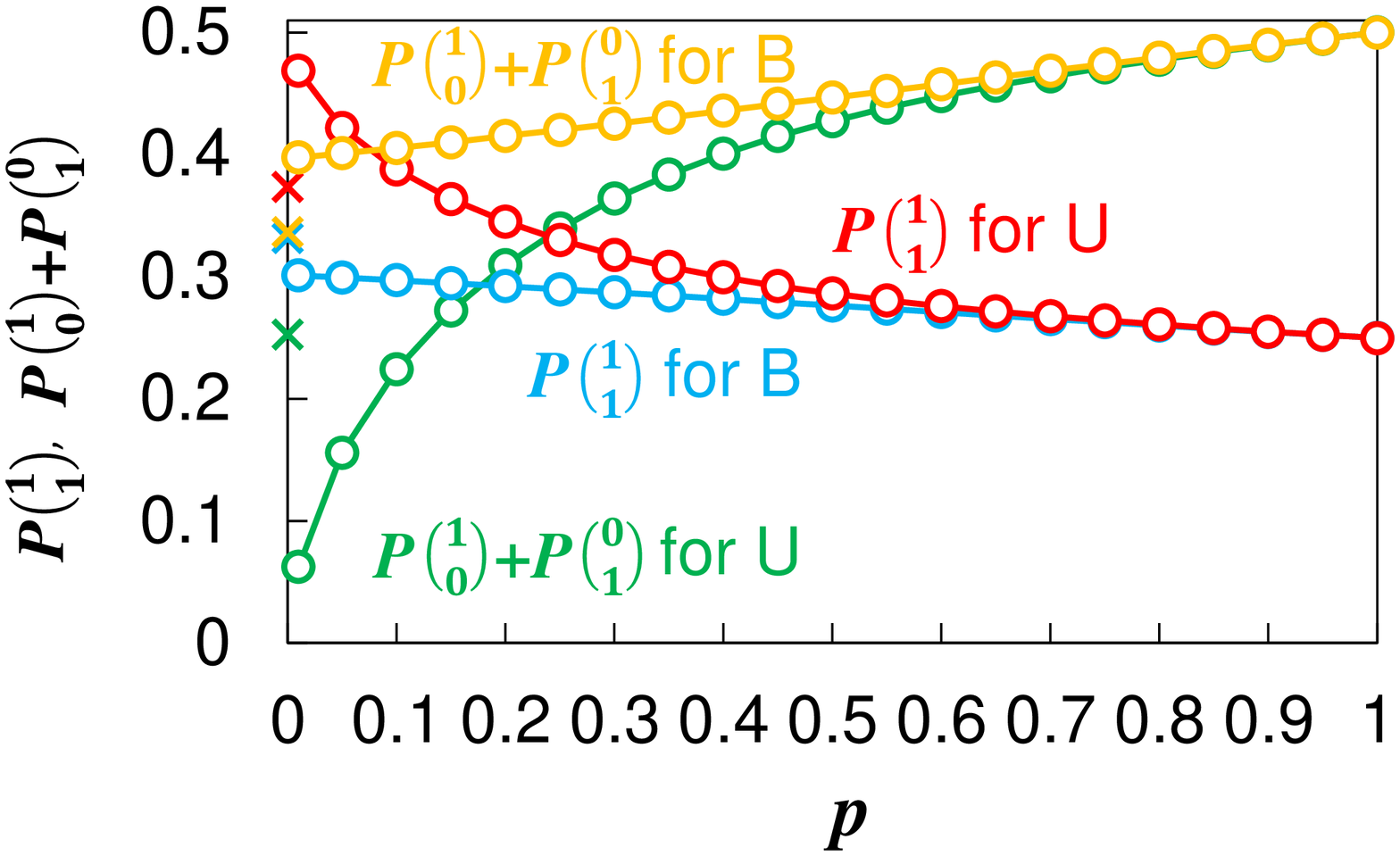}
\caption{(Color Online) Simulation results for steady-state values of $P(\begin{smallmatrix} 1 \\ 0 \end{smallmatrix})+P(\begin{smallmatrix} 0 \\ 1 \end{smallmatrix})$, and $P(\begin{smallmatrix} 1 \\ 1 \end{smallmatrix})$ as functions of $p$, fixing $(\omega_{\rm A1},\omega_{\rm A2},\omega_{\rm D1},\omega_{\rm D2})=(0,0,0,0)$ and $\rho=0.5$. $P(\begin{smallmatrix} 0 \\ 0 \end{smallmatrix})=P(\begin{smallmatrix} 1 \\ 1 \end{smallmatrix})$ when $\rho=0.5$. The points between $p=0$ and $p=0.05$ are for $p=0.01$. We note (i) that the values for $p=0$ are calculated by averaging 100 different initial configurations and (ii) that they are calculated steps for $10^9$ time steps after evolving the system with $p=0.01$ for $10^9$ time steps  because it takes more time for the system to evolve into the steady state.}
\label{fig:P1P2P3}
\end{center}
\end{figure}

First, to investigate the influence of $p$ on $\rho$, we investigate $P(\begin{smallmatrix} 0 \\ 0 \end{smallmatrix})$, $P(\begin{smallmatrix} 1 \\ 0 \end{smallmatrix})+P(\begin{smallmatrix} 0 \\ 1 \end{smallmatrix})$, and $P(\begin{smallmatrix} 1 \\ 1 \end{smallmatrix})$, fixing $(\omega_{\rm A1},\omega_{\rm A2},\omega_{\rm D1},\omega_{\rm D2})=(0,0,0,0)$ and $\rho=0.5$, as in Fig. \ref{fig:P1P2P3}.

Figure \ref{fig:P1P2P3} shows (i) that $P(\begin{smallmatrix} 1 \\ 1 \end{smallmatrix})$ and $P(\begin{smallmatrix} 0 \\ 0 \end{smallmatrix})$ increase and $P(\begin{smallmatrix} 1 \\ 0 \end{smallmatrix})+P(\begin{smallmatrix} 0 \\ 1 \end{smallmatrix})$ decreases with smaller $p$ for the two directional flows, (ii) that the degrees of those changes are greater for unidirectional flows than for bidirectional flows, excluding the case with $p=0$ for unidirectional flows, and (iii) that $P(\begin{smallmatrix} 1 \\ 1 \end{smallmatrix})$ and $P(\begin{smallmatrix} 0 \\ 0 \end{smallmatrix})$ decrease and $P(\begin{smallmatrix} 1 \\ 0 \end{smallmatrix})+P(\begin{smallmatrix} 0 \\ 1 \end{smallmatrix})$ increases abruptly from $p=0.01$ to $p=0$ for unidirectional flows.

\begin{figure}[htbp]
\begin{center}
\includegraphics[width=8.5cm,clip]{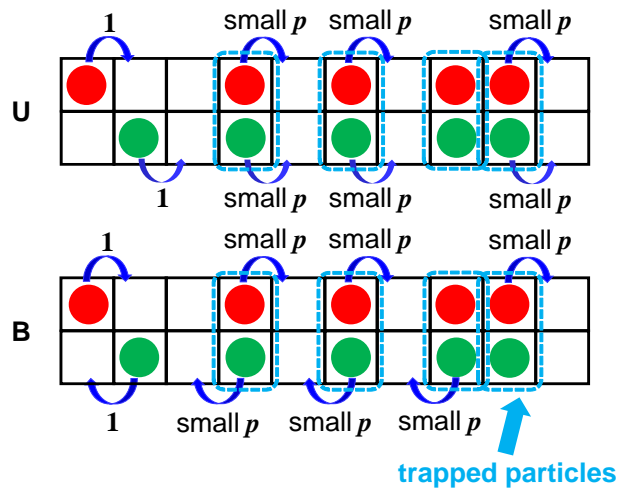}
\caption{(Color Online) Schematic illustration of the trapping effect. Langmuir kinetics are not considered in this figure, which shows the case $L=10$.}
\label{fig:trap}
\end{center}
\end{figure}

Three effects can qualitatively explain these phenomena: (i) the trapping, (ii) jamming, and (iii) blocking effects. First, in the trapping effect, for small $p$, particles can become trapped in the state $(\begin{smallmatrix} 1 \\ 1 \end{smallmatrix})$, which leads to an increase in $P(\begin{smallmatrix} 1 \\ 1 \end{smallmatrix})$ and $P(\begin{smallmatrix} 0 \\ 0 \end{smallmatrix})$, and a decrease in $P(\begin{smallmatrix} 1 \\ 0 \end{smallmatrix})+P(\begin{smallmatrix} 0 \\ 1 \end{smallmatrix})$, as shown in Fig. \ref{fig:trap}. We stress that the trapping effect works in common between the two directional flows. 

\begin{figure}[htbp]
\begin{center}
\includegraphics[width=8.5cm,clip]{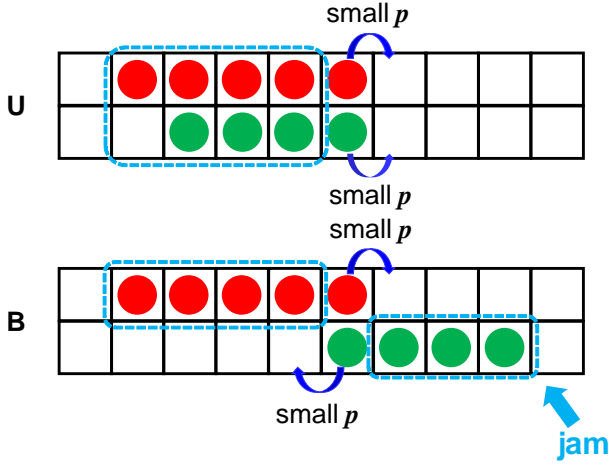}
\caption{(Color Online) Schematic illustration of the jamming effect. Langmuir kinetics are not considered in this figure, which shows the case $L=10$.}
\label{fig:jam}
\end{center}
\end{figure}

Contrarily, in the jamming effect, for small $p$, the particles following a trapped particle can become involved in a jam, as shown in Fig. \ref{fig:jam}. For unidirectional flows, this effect works the same as the trapping effect; specifically, $P(\begin{smallmatrix} 1 \\ 1 \end{smallmatrix})$ and $P(\begin{smallmatrix} 0 \\ 0 \end{smallmatrix})$ increase and $P(\begin{smallmatrix} 1 \\ 0 \end{smallmatrix})+P(\begin{smallmatrix} 0 \\ 1 \end{smallmatrix})$ decreases with smaller $p$. In contrast, for bidirectional flows, this effect tends to counter the trapping effect; specifically, $P(\begin{smallmatrix} 1 \\ 1 \end{smallmatrix})$ and $P(\begin{smallmatrix} 0 \\ 0 \end{smallmatrix})$ decrease, and $P(\begin{smallmatrix} 1 \\ 0 \end{smallmatrix})+P(\begin{smallmatrix} 0 \\ 1 \end{smallmatrix})$ increases with smaller $p$. We note that the jamming effect is smaller than the trapping effect because jams occur only after many trapped particles appear. Because the jamming effect influences each $P$ oppositely in the two directional flows, the degrees of those changes are greater for unidirectional flows than for bidirectional flows, excluding the case with $p=0$.

\begin{figure}[htbp]
\begin{center}
\includegraphics[width=8.5cm,clip]{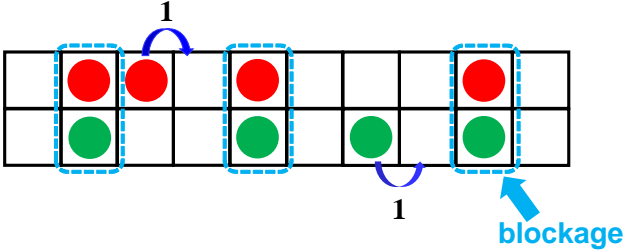}
\caption{(Color Online) Schematic illustration of the blocking effect for unidirectional flows. Langmuir kinetics are not considered in this figure, which shows the case $L=10$.}
\label{fig:block}
\end{center}
\end{figure}

Finally, the blocking effect appears only for $p=0$, as shown in Fig. \ref{fig:block}. Once the particles are trapped in the state $(\begin{smallmatrix} 1 \\ 1 \end{smallmatrix})$, this state never changes for $p=0$ and $\omega_{\rm D2}=0$; therefore, the isolated particles between the two clusters of $(\begin{smallmatrix} 1 \\ 1 \end{smallmatrix})$ must finally make the state $(\begin{smallmatrix} 1 \\ 0 \end{smallmatrix})$ or $(\begin{smallmatrix} 0 \\ 1 \end{smallmatrix})$. Because this effect counters the trapping and jamming effects, $P(\begin{smallmatrix} 1 \\ 1 \end{smallmatrix})$ and $P(\begin{smallmatrix} 0 \\ 0 \end{smallmatrix})$ decrease and $P(\begin{smallmatrix} 1 \\ 0 \end{smallmatrix})+P(\begin{smallmatrix} 0 \\ 1 \end{smallmatrix})$ increases from $p=0.01$ to $p=0$ for unidirectional flows. We note (i) that this effect is virtually the same as the jamming effect for bidirectional flows and (ii) that the blockages can be dismantled if $\omega_{\rm D2}>0$.

We can also confirm the existence of those three effects in Fig. \ref{fig:mu}. Table \ref{tab:mu} summarizes the kinds of $\mu$, which become larger with smaller $p$. We note that Table \ref{tab:mu} excludes the case $p=0$ for unidirectional flows because of its singularity.

For unidirectional flows, the trapping and jamming effects explain why all the listed $\mu$ values become larger with smaller $p$. Contrarily, for bidirectional flows, the fact that $\mu(\begin{smallmatrix} 0 & 0 \\ 0 & 0 \end{smallmatrix})$, $\mu(\begin{smallmatrix} 1 & 0 \\ 1 & 1 \end{smallmatrix})$, and $\mu(\begin{smallmatrix} 1 & 1 \\ 0 & 1 \end{smallmatrix})$ become larger with smaller $p$ can be explained by the trapping effect, whereas the fact that $\mu(\begin{smallmatrix} 1 & 1 \\ 0 & 0 \end{smallmatrix})$ and $\mu(\begin{smallmatrix} 0 & 0 \\ 1 & 1 \end{smallmatrix})$ become larger with smaller $p$ can be explained by the jamming effect. Conversely, for $p=0$, $\mu(\begin{smallmatrix} 1 & 1 \\ 0 & 0 \end{smallmatrix})$, $\mu(\begin{smallmatrix} 1 & 1 \\ 0 & 1 \end{smallmatrix})$, $\mu(\begin{smallmatrix} 0 & 0 \\ 1 & 1 \end{smallmatrix})$, and $\mu(\begin{smallmatrix} 1 & 0 \\ 1 & 1 \end{smallmatrix})$ also soar for unidirectional flows, indicating the blocking effect.

\begin{table}[htbp]
\centering
\caption{Kinds of $\mu$, which become larger with smaller $p$, for each direction (see Fig. \ref{fig:mu}).}
\label{tab:mu}
\begin{tabular}{c|c}
Direction & Corresponding $\mu$ \\ \hline \hline
Unidirection & $\mu(\begin{smallmatrix} 0 & 0 \\ 0 & 0 \end{smallmatrix})$, $\mu(\begin{smallmatrix} 1 & 1 \\ 0 & 1 \end{smallmatrix})$, $\mu(\begin{smallmatrix} 0 & 1 \\ 1 & 1 \end{smallmatrix})$, $\mu(\begin{smallmatrix} 1 & 1 \\ 1 & 1 \end{smallmatrix})$ \\ \hline
Bidirection & $\mu(\begin{smallmatrix} 0 & 0 \\ 0 & 0 \end{smallmatrix})$, $\mu(\begin{smallmatrix} 1 & 1 \\ 0 & 0 \end{smallmatrix})$, $\mu(\begin{smallmatrix} 1 & 1 \\ 0 & 1 \end{smallmatrix})$, $\mu(\begin{smallmatrix} 0 & 0 \\ 1 & 1 \end{smallmatrix})$, $\mu(\begin{smallmatrix} 1 & 0 \\ 1 & 1 \end{smallmatrix})$\\ 
\end{tabular}
\end{table} 

On the basis aforementioned discussions, we finally summarize the influences of the three effects on $P(\begin{smallmatrix} 0 \\ 0 \end{smallmatrix})$, $P(\begin{smallmatrix} 1 \\ 0 \end{smallmatrix})+P(\begin{smallmatrix} 0 \\ 1 \end{smallmatrix})$, and $P(\begin{smallmatrix} 1 \\ 1 \end{smallmatrix})$ in Tab. \ref{tab:3effect}. We again stress that the results shown in Fig. \ref{fig:P1P2P3} can be explained from Tab. \ref{tab:3effect}, noting that the trapping effect is stronger than the jamming effect and that the blocking effect only appears for $p=0$.

\begin{table}[htbp]
\centering
\caption{Influence of the three effects on $P(\begin{smallmatrix} 0 \\ 0 \end{smallmatrix})$, $P(\begin{smallmatrix} 1 \\ 0 \end{smallmatrix})+P(\begin{smallmatrix} 0 \\ 1 \end{smallmatrix})$, and $P(\begin{smallmatrix} 1 \\ 1 \end{smallmatrix})$. The sign “$+$” (“$-$”) indicates that the corresponding effect causes increases (decreases) in each $P$. Langmuir kinetics are not considered in this table. We note that “$++$” (“$--$”), used for unidirectional flows, represents the fact that the effect is larger than that for bidirectional flows.}
\label{tab:3effect}
\begin{tabular}{c|c|c|c|c}
Direction & Effect & $P(\begin{smallmatrix} 0 \\ 0 \end{smallmatrix})$ & $P(\begin{smallmatrix} 1 \\ 0 \end{smallmatrix})+P(\begin{smallmatrix} 0 \\ 1 \end{smallmatrix})$ & $P(\begin{smallmatrix} 1 \\ 1 \end{smallmatrix})$ \\ \hline \hline
\multirow{4}{*}{Unidirection} & Trapping effect & $+$ & $-$ & $+$ \\ \cline{2-5}
& Jamming effect & $+$ & $-$ & $+$ \\ \cline{2-5}
& Blocking effect& $-$ & $+$ & $-$ \\ \cline{2-5}
& Trapping $+$ Jamming& $++$ & $--$ & $++$ \\ \hline
\multirow{4}{*}{Bidirection} & Trapping effect & $+$ & $-$ & $+$ \\ \cline{2-5}
& Jamming effect & $-$ & $+$ & $-$ \\ \cline{2-5}
& Blocking effect & $-$ & $+$ & $-$ \\ \cline{2-5}
& Trapping$+$Jamming& $+$ & $-$ & $+$ \\ \hline
\end{tabular}
\end{table}

Next, we consider the influences of $\omega$ on $\rho$, $P(\begin{smallmatrix} 0 \\ 0 \end{smallmatrix})$, $P(\begin{smallmatrix} 1 \\ 0 \end{smallmatrix})+P(\begin{smallmatrix} 0 \\ 1 \end{smallmatrix})$, and $P(\begin{smallmatrix} 1 \\ 1 \end{smallmatrix})$. 



\begin{figure}[htbp]
\begin{center}
\includegraphics[width=7cm,clip]{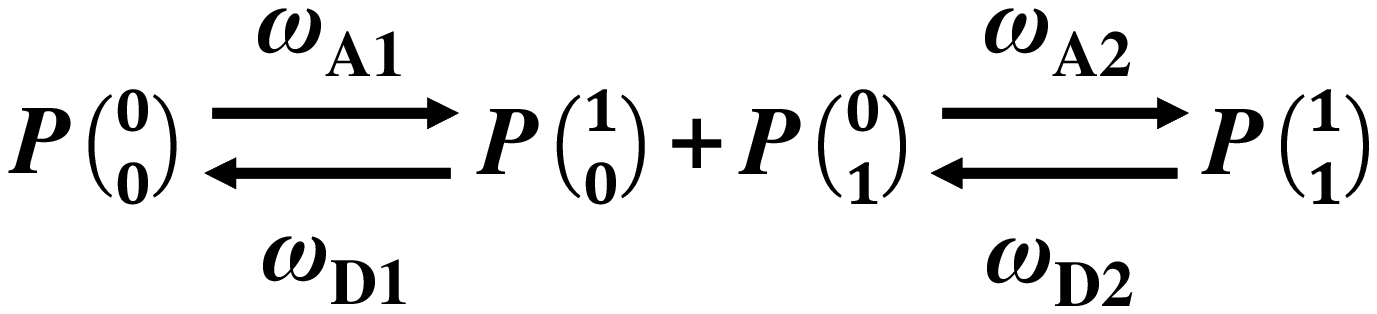}
\caption{State-transition diagram for $P(\begin{smallmatrix} 0 \\ 0 \end{smallmatrix})$, $P(\begin{smallmatrix} 1 \\ 0 \end{smallmatrix})+P(\begin{smallmatrix} 0 \\ 1 \end{smallmatrix})$, and $P(\begin{smallmatrix} 1 \\ 1 \end{smallmatrix})$.}
\label{fig:transition}
\end{center}
\end{figure}

From the definition of $\omega$ and the state-transition of diagram, as shown in Fig. \ref{fig:transition}, we can summarize the increase ($+$) and decrease ($-$) of each value by the change of $\omega$ in Tab. \ref{tab:omega}. Blank cells indicate that the effect is indirect and the sign is not apparent; however, those indirect effects are small enough to be ignored in the following discussion.  

\begin{table}[htbp]
\centering
\caption{Increase ($+$) and decrease ($-$) of each value caused by the change in $\omega$ ($\Delta \omega>0$). For example, the first row in the column for $\rho$ indicates that $\rho$ increases with larger $\omega_{\rm A1}$.}
\label{tab:omega}
\begin{tabular}{c|c|c|c|c}
Change of $\omega$ & $\rho$ & $P(\begin{smallmatrix} 0 \\ 0 \end{smallmatrix})$ & $P(\begin{smallmatrix} 1 \\ 0 \end{smallmatrix})+P(\begin{smallmatrix} 0 \\ 1 \end{smallmatrix})$ & $P(\begin{smallmatrix} 1 \\ 1 \end{smallmatrix})$ \\ \hline \hline
$\omega_{\rm A1}\to\omega_{\rm A1}+\Delta \omega$ & $+$ & $-$ & $+$ & \\ \hline
$\omega_{\rm A2}\to\omega_{\rm A2}+\Delta \omega$ & $+$ &  & $-$ & $+$\\ \hline
$\omega_{\rm D1}\to\omega_{\rm D1}+\Delta \omega$ & $-$ & $+$ & $-$ & \\ \hline
$\omega_{\rm D2}\to\omega_{\rm D2}+\Delta \omega$ & $-$ &  & $+$ & $-$
\end{tabular}
\end{table}

On the basis of the aforementioned discussions of $(p,\omega)$-dependence of $\rho$, we consider the following four phenomena in Fig. \ref{fig:general}. We stress that those phenomena can be confirmed not only for the simulation results but also for the numerical results with 4-cluster mean-field analysis.

\paragraph{Increase/decrease of $\rho$, depending on $\omega$, in the region where $p\gtrapprox0.1$}

--- Because $P(\begin{smallmatrix} 0 \\ 0 \end{smallmatrix})$ and $P(\begin{smallmatrix} 1 \\ 1 \end{smallmatrix})$ increase and $P(\begin{smallmatrix} 1 \\ 0 \end{smallmatrix})+P(\begin{smallmatrix} 0 \\ 1 \end{smallmatrix})$ decreases for smaller $p$, the effects of $\omega_{\rm A1}$ and $\omega_{\rm D2}$ ($\omega_{\rm A2}$ and $\omega_{\rm D1}$) become smaller. Therefore, considering Fig. \ref{fig:transition} and Tab. \ref{tab:omega}, $\rho$ increases (decreases) for smaller $p$ for Cases (a), (d), (e), and (h) [(b), (c), (f), and (g)] in the region where $p\gtrapprox0.1$. For example, for Case (a), $P(\begin{smallmatrix} 0 \\ 0 \end{smallmatrix})$ increases, and the effect of $\omega_{\rm A1}$ is enhanced for smaller $p$, resulting in an increase in $\rho$ for smaller $p$.

\paragraph{Differences in the degree of change between the two directional flows}

--- Because the effect of $p$ on each $P$ is larger for unidirectional flows than for bidirectional ones (see Fig. \ref{fig:P1P2P3} and Tab. \ref{tab:3effect}), the degree of the change in $\rho$ with $p$ becomes greater for unidirectional flows than for bidirectional ones.

\paragraph{Change of the trend in unidirectional flows for $p=0$}

--- For unidirectional flows, due to the blocking effect, the configuration changes drastically from very small $p(\neq0)$ to $p=0$, although Langmuir kinetics reduce that effect. This results in a trend change when $p=0$, the extent of which depends on $\omega$. We note that for unidirectional flows the influence of $p$ on $P$ is large enough so that it is not necessary to consider the influence of $\omega$. 

\paragraph{Change of the trend in bidirectional flows when $p$ becomes smaller} 

--- Unlike unidirectional flows, we cannot ignore the effect of $\omega$ on $P$ for bidirectional flows. Therefore, for Cases (a), (d), (f), and (g), where the influence of $\omega$ on $P$ counters that of smaller $p$, the trend changes in the change-easing direction with smaller $p$. In contrast, for Cases (b), (c), (e), and (h), where the influence of $\omega$ on $P$ reinforces that of smaller $p$, the trend changes in the change-accelerating direction with smaller $p$.

\section{CONCLUSION}
\label{sec:conclusion}

In the present paper, we have investigated a two-lane extended LK-TASEP on a periodic lattice, where the hopping rate $p$ and the attachment (detachment) rate $\omega_{\rm A}$ ($\omega_{\rm D}$) vary depending on the state of the corresponding site in the other lane. The proposed model is new in that it introduces a varying rule for the attachment and detachment rate. We have investigated the steady-state global density for unidirectional and bidirectional flows using both computer simulations and mean-field analyses.

We have conducted three kinds of mean-field analyses (1-, 2-, and 4-cluster cases) and have compared them with simulation results. In the 1-cluster mean-field analysis, the calculated value of $\rho$ is in good agreement with the simulation results only in the region where $p$ is near 1. In contrast, the 2-cluster analysis can reproduce the rough trend of the simulation results for $\rho$ as functions of $p$, even though it cannot distinguish between unidirectional and bidirectional flows. Finally, the 4-cluster analysis can not only approximates better the simulation results for $\rho$ as functions of $p$ but also reproduces the difference between the two directional flows.

We have therefore considered further the discrepancies between the numerical results from 2- and 4-cluster mean-field analyses for unidirectional flows---which are smaller than those for bidirectional flows---by calculating the correlations between two adjacent ($2\times1$) clusters. We have also discussed the $(p,\omega)$-dependence of $\rho$ in terms of three effects (the trapping, jamming, and blocking effect). Those three effects by $p$ and $\omega$ determine the trend of $\rho$. 

We again emphasize that, despite its simplicity, the proposed model has a potential for applications to real-world phenomena. For example, for crowd dynamics (traffic flow) in a narrow passage (road), the proposed model can consider the velocity and inflow/outflow of pedestrians (vehicles). In particular, unlike previous models, our model makes it possible to consider of the dependence of the changes of the inflow/outflow on the lane state.

\section*{ACKNOWLEDGMENTS}
This work was partially supported by JST-Mirai Program Grant Number JPMJMI17D4, Japan, JSPS KAKENHI Grant Number JP15K17583.

\appendix

\section{Exclusion of the other solution of Eq. (\ref{eq:original4})}
\label{sec:1clusterd}
In this Appendix, we discuss the exclusion of the other solution of Eq. (\ref{eq:original4}); specifically, 
\begin{equation}
\rho=\frac{-b+\sqrt{b^2-4ac}}{2a}.
\end{equation}

First, we consider the case where $a>0$; specifically,
\begin{equation}
\omega_{\rm A1}+\omega_{\rm D1}>\omega_{\rm A2}+\omega_{\rm D2}.
\end{equation}
In this case, we have
\begin{eqnarray}
b&=&-2\omega_{\rm A1}+\omega_{\rm A2}-\omega_{\rm D1}\\
&=&-(\omega_{\rm A1}+\omega_{\rm D1})-\omega_{\rm A1}+\omega_{\rm A2}\\
&<&-(\omega_{\rm A2}+\omega_{\rm D2})-\omega_{\rm A1}+\omega_{\rm A2}\\
&=&-\omega_{\rm A1}-\omega_{\rm D2},
\end{eqnarray}
resulting in $b<0$.
Noting that $a>0$, $b<0$ and $c>0$, we have
\begin{eqnarray}
&&\frac{-b+\sqrt{b^2-4ac}}{2a}\\
&&>\frac{-b-b}{2a}=-\frac{b}{a}\\
&&=\frac{2\omega_{\rm A1}-\omega_{\rm A2}+\omega_{\rm D1}}{\omega_{\rm A1}-\omega_{\rm A2}+\omega_{\rm D1}-\omega_{\rm D2}}\\
&&=\frac{\omega_{\rm A1}-\omega_{\rm A2}+\omega_{\rm D1}+\omega_{\rm A1}}{\omega_{\rm A1}-\omega_{\rm A2}+\omega_{\rm D1}-\omega_{\rm D2}}>1.
\end{eqnarray}

Second, we consider the case where $a<0$. In this case, because
\begin{equation}
-b+\sqrt{b^2-4ac}>-b+|b|\geq0,
\end{equation}
we have
\begin{equation}
\frac{-b+\sqrt{b^2-4ac}}{2a}<0.
\end{equation}

Because $\rho$ lies in the range $0\leq\rho\leq1$, this solution is inappropriate, and because the system must evolve into only one steady state, we obtain Eq. (\ref{eq:original5}).

\section{Exclusion of the other solution of Eq. (\ref{eq:P1eq})}
\label{sec:2clusterd}
In this Appendix, we discuss the exclusion of the other solution of Eq. (\ref{eq:P1eq}); specifically, 
\begin{equation}
P(
)$ lies in the range $0\leq P(%
)\leq1$, this solution is inappropriate, and because the system must evolve into only one steady state, we obtain Eq. (\ref{eq:P1}).

\begin{widetext}
\section{Other independent master equations for the 4-cluster mean-field analysis for unidirectional flows}
\label{sec:4othermasteruni}

In this Appendix, we summarize the other independent master equations for 4-cluster probabilities for unidirectional flows. 

For $P(%
)$, the master equations can be expressed as

\begin{equation}
\begin{split}
\frac{d P(%
)\bigr].\\
\end{split}
\label{eq:4cluster13}
\end{equation}

Inserting Eq. (\ref{eq:condition}) into Eqs. (\ref{eq:4cluster30})--(\ref{eq:4cluster13}) and noting that in the steady state, $\frac{d}{dt}P(%
)=0,\\
\end{split}
\label{eq:4cluster132}
\end{equation}
respectively.

\section{Other independent master equations for the 4-cluster mean-field analysis for bidirectional flows}
\label{sec:4othermasterbi}

In this Appendix, we summarize the other independent master equations for 4-cluster probabilities for bidirectional flows. 

For $P(%
)$, the master equations can be expressed as

\begin{equation}
\begin{split}
\frac{d P(%
)\bigr].\\
\end{split}
\label{eq:4cluster32b}
\end{equation}

Inserting Eq. (\ref{eq:condition}) into Eqs. (\ref{eq:4cluster12b})--(\ref{eq:4cluster32b}) and noting that in the steady state $\frac{d}{dt}P(%
)=0,\\
\end{split}
\label{eq:4cluster322b}
\end{equation}
respectively.

\end{widetext}

\appendix{}


\begin{thebibliography}{36}%
\makeatletter
\providecommand \@ifxundefined [1]{%
 \@ifx{#1\undefined}
}%
\providecommand \@ifnum [1]{%
 \ifnum #1\expandafter \@firstoftwo
 \else \expandafter \@secondoftwo
 \fi
}%
\providecommand \@ifx [1]{%
 \ifx #1\expandafter \@firstoftwo
 \else \expandafter \@secondoftwo
 \fi
}%
\providecommand \natexlab [1]{#1}%
\providecommand \enquote  [1]{``#1''}%
\providecommand \bibnamefont  [1]{#1}%
\providecommand \bibfnamefont [1]{#1}%
\providecommand \citenamefont [1]{#1}%
\providecommand \href@noop [0]{\@secondoftwo}%
\providecommand \href [0]{\begingroup \@sanitize@url \@href}%
\providecommand \@href[1]{\@@startlink{#1}\@@href}%
\providecommand \@@href[1]{\endgroup#1\@@endlink}%
\providecommand \@sanitize@url [0]{\catcode `\\12\catcode `\$12\catcode
  `\&12\catcode `\#12\catcode `\^12\catcode `\_12\catcode `\%12\relax}%
\providecommand \@@startlink[1]{}%
\providecommand \@@endlink[0]{}%
\providecommand \url  [0]{\begingroup\@sanitize@url \@url }%
\providecommand \@url [1]{\endgroup\@href {#1}{\urlprefix }}%
\providecommand \urlprefix  [0]{URL }%
\providecommand \Eprint [0]{\href }%
\providecommand \doibase [0]{http://dx.doi.org/}%
\providecommand \selectlanguage [0]{\@gobble}%
\providecommand \bibinfo  [0]{\@secondoftwo}%
\providecommand \bibfield  [0]{\@secondoftwo}%
\providecommand \translation [1]{[#1]}%
\providecommand \BibitemOpen [0]{}%
\providecommand \bibitemStop [0]{}%
\providecommand \bibitemNoStop [0]{.\EOS\space}%
\providecommand \EOS [0]{\spacefactor3000\relax}%
\providecommand \BibitemShut  [1]{\csname bibitem#1\endcsname}%
\let\auto@bib@innerbib\@empty
\bibitem [{\citenamefont {Schadschneider}\ \emph {et~al.}(2010)\citenamefont
  {Schadschneider}, \citenamefont {Chowdhury},\ and\ \citenamefont
  {Nishinari}}]{schadschneider2010stochastic}%
  \BibitemOpen
  \bibfield  {author} {\bibinfo {author} {\bibfnamefont {A.}~\bibnamefont
  {Schadschneider}}, \bibinfo {author} {\bibfnamefont {D.}~\bibnamefont
  {Chowdhury}}, \ and\ \bibinfo {author} {\bibfnamefont {K.}~\bibnamefont
  {Nishinari}},\ }\href@noop {} {\emph {\bibinfo {title} {Stochastic transport
  in complex systems: from molecules to vehicles}}}\ (\bibinfo  {publisher}
  {Elsevier},\ \bibinfo {year} {2010})\BibitemShut {NoStop}%
\bibitem [{\citenamefont {MacDonald}\ \emph {et~al.}(1968)\citenamefont
  {MacDonald}, \citenamefont {Gibbs},\ and\ \citenamefont
  {Pipkin}}]{macdonald1968kinetics}%
  \BibitemOpen
  \bibfield  {author} {\bibinfo {author} {\bibfnamefont {C.~T.}\ \bibnamefont
  {MacDonald}}, \bibinfo {author} {\bibfnamefont {J.~H.}\ \bibnamefont
  {Gibbs}}, \ and\ \bibinfo {author} {\bibfnamefont {A.~C.}\ \bibnamefont
  {Pipkin}},\ }\href@noop {} {\bibfield  {journal} {\bibinfo  {journal}
  {Biopolymers}\ }\textbf {\bibinfo {volume} {6}},\ \bibinfo {pages} {1}
  (\bibinfo {year} {1968})}\BibitemShut {NoStop}%
\bibitem [{\citenamefont {MacDonald}\ and\ \citenamefont
  {Gibbs}(1969)}]{macdonald1969concerning}%
  \BibitemOpen
  \bibfield  {author} {\bibinfo {author} {\bibfnamefont {C.~T.}\ \bibnamefont
  {MacDonald}}\ and\ \bibinfo {author} {\bibfnamefont {J.~H.}\ \bibnamefont
  {Gibbs}},\ }\href@noop {} {\bibfield  {journal} {\bibinfo  {journal}
  {Biopolymers}\ }\textbf {\bibinfo {volume} {7}},\ \bibinfo {pages} {707}
  (\bibinfo {year} {1969})}\BibitemShut {NoStop}%
\bibitem [{\citenamefont {Chou}\ \emph {et~al.}(2011)\citenamefont {Chou},
  \citenamefont {Mallick},\ and\ \citenamefont {Zia}}]{chou2011non}%
  \BibitemOpen
  \bibfield  {author} {\bibinfo {author} {\bibfnamefont {T.}~\bibnamefont
  {Chou}}, \bibinfo {author} {\bibfnamefont {K.}~\bibnamefont {Mallick}}, \
  and\ \bibinfo {author} {\bibfnamefont {R.}~\bibnamefont {Zia}},\ }\href@noop
  {} {\bibfield  {journal} {\bibinfo  {journal} {Rep. Prog. Phys.}\ }\textbf
  {\bibinfo {volume} {74}},\ \bibinfo {pages} {116601} (\bibinfo {year}
  {2011})}\BibitemShut {NoStop}%
\bibitem [{\citenamefont {Appert-Rolland}\ \emph {et~al.}(2015)\citenamefont
  {Appert-Rolland}, \citenamefont {Ebbinghaus},\ and\ \citenamefont
  {Santen}}]{appert2015intracellular}%
  \BibitemOpen
  \bibfield  {author} {\bibinfo {author} {\bibfnamefont {C.}~\bibnamefont
  {Appert-Rolland}}, \bibinfo {author} {\bibfnamefont {M.}~\bibnamefont
  {Ebbinghaus}}, \ and\ \bibinfo {author} {\bibfnamefont {L.}~\bibnamefont
  {Santen}},\ }\href@noop {} {\bibfield  {journal} {\bibinfo  {journal} {Phys.
  Rep.}\ }\textbf {\bibinfo {volume} {593}},\ \bibinfo {pages} {1} (\bibinfo
  {year} {2015})}\BibitemShut {NoStop}%
\bibitem [{\citenamefont {Ghosh}\ \emph {et~al.}(2019)\citenamefont {Ghosh},
  \citenamefont {Dutta}, \citenamefont {Patra}, \citenamefont {Sato},
  \citenamefont {Nishinari},\ and\ \citenamefont
  {Chowdhury}}]{PhysRevE.99.052122}%
  \BibitemOpen
  \bibfield  {author} {\bibinfo {author} {\bibfnamefont {S.}~\bibnamefont
  {Ghosh}}, \bibinfo {author} {\bibfnamefont {A.}~\bibnamefont {Dutta}},
  \bibinfo {author} {\bibfnamefont {S.}~\bibnamefont {Patra}}, \bibinfo
  {author} {\bibfnamefont {J.}~\bibnamefont {Sato}}, \bibinfo {author}
  {\bibfnamefont {K.}~\bibnamefont {Nishinari}}, \ and\ \bibinfo {author}
  {\bibfnamefont {D.}~\bibnamefont {Chowdhury}},\ }\href {\doibase
  10.1103/PhysRevE.99.052122} {\bibfield  {journal} {\bibinfo  {journal} {Phys.
  Rev. E}\ }\textbf {\bibinfo {volume} {99}},\ \bibinfo {pages} {052122}
  (\bibinfo {year} {2019})}\BibitemShut {NoStop}%
\bibitem [{\citenamefont {Zarai}\ \emph {et~al.}(2017)\citenamefont {Zarai},
  \citenamefont {Margaliot},\ and\ \citenamefont
  {Tuller}}]{10.1371/journal.pone.0182178}%
  \BibitemOpen
  \bibfield  {author} {\bibinfo {author} {\bibfnamefont {Y.}~\bibnamefont
  {Zarai}}, \bibinfo {author} {\bibfnamefont {M.}~\bibnamefont {Margaliot}}, \
  and\ \bibinfo {author} {\bibfnamefont {T.}~\bibnamefont {Tuller}},\ }\href
  {\doibase 10.1371/journal.pone.0182178} {\bibfield  {journal} {\bibinfo
  {journal} {PLOS ONE}\ }\textbf {\bibinfo {volume} {12}},\ \bibinfo {pages}
  {1} (\bibinfo {year} {2017})}\BibitemShut {NoStop}%
\bibitem [{\citenamefont {Helbing}(2001)}]{RevModPhys.73.1067}%
  \BibitemOpen
  \bibfield  {author} {\bibinfo {author} {\bibfnamefont {D.}~\bibnamefont
  {Helbing}},\ }\href {\doibase 10.1103/RevModPhys.73.1067} {\bibfield
  {journal} {\bibinfo  {journal} {Rev. Mod. Phys.}\ }\textbf {\bibinfo {volume}
  {73}},\ \bibinfo {pages} {1067} (\bibinfo {year} {2001})}\BibitemShut
  {NoStop}%
\bibitem [{\citenamefont {Yamamoto}\ \emph {et~al.}(2017)\citenamefont
  {Yamamoto}, \citenamefont {Yanagisawa},\ and\ \citenamefont
  {Nishinari}}]{yamamoto2017velocity}%
  \BibitemOpen
  \bibfield  {author} {\bibinfo {author} {\bibfnamefont {H.}~\bibnamefont
  {Yamamoto}}, \bibinfo {author} {\bibfnamefont {D.}~\bibnamefont
  {Yanagisawa}}, \ and\ \bibinfo {author} {\bibfnamefont {K.}~\bibnamefont
  {Nishinari}},\ }\href@noop {} {\bibfield  {journal} {\bibinfo  {journal} {J.
  Stat. Mech.}\ }\textbf {\bibinfo {volume} {2017}},\ \bibinfo {pages} {043204}
  (\bibinfo {year} {2017})}\BibitemShut {NoStop}%
\bibitem [{\citenamefont {Arita}\ and\ \citenamefont
  {Schadschneider}(2015)}]{arita2015exclusive}%
  \BibitemOpen
  \bibfield  {author} {\bibinfo {author} {\bibfnamefont {C.}~\bibnamefont
  {Arita}}\ and\ \bibinfo {author} {\bibfnamefont {A.}~\bibnamefont
  {Schadschneider}},\ }\href@noop {} {\bibfield  {journal} {\bibinfo  {journal}
  {Math. Mod Meth.}\ }\textbf {\bibinfo {volume} {25}},\ \bibinfo {pages} {401}
  (\bibinfo {year} {2015})}\BibitemShut {NoStop}%
\bibitem [{\citenamefont {Yamamoto}\ \emph {et~al.}(2019)\citenamefont
  {Yamamoto}, \citenamefont {Yanagisawa},\ and\ \citenamefont
  {Nishinari}}]{PhysRevE.100.042106}%
  \BibitemOpen
  \bibfield  {author} {\bibinfo {author} {\bibfnamefont {H.}~\bibnamefont
  {Yamamoto}}, \bibinfo {author} {\bibfnamefont {D.}~\bibnamefont
  {Yanagisawa}}, \ and\ \bibinfo {author} {\bibfnamefont {K.}~\bibnamefont
  {Nishinari}},\ }\href {\doibase 10.1103/PhysRevE.100.042106} {\bibfield
  {journal} {\bibinfo  {journal} {Phys. Rev. E}\ }\textbf {\bibinfo {volume}
  {100}},\ \bibinfo {pages} {042106} (\bibinfo {year} {2019})}\BibitemShut
  {NoStop}%
\bibitem [{\citenamefont {Yanagisawa}(2016)}]{CDA8}%
  \BibitemOpen
  \bibfield  {author} {\bibinfo {author} {\bibfnamefont {D.}~\bibnamefont
  {Yanagisawa}},\ }\href {\doibase 10.17815/CD.2016.8} {\bibfield  {journal}
  {\bibinfo  {journal} {Collect. Dyn.}\ }\textbf {\bibinfo {volume} {1}},\
  \bibinfo {pages} {1} (\bibinfo {year} {2016})}\BibitemShut {NoStop}%
\bibitem [{\citenamefont {Hao}\ \emph {et~al.}(2010)\citenamefont {Hao},
  \citenamefont {Jiang}, \citenamefont {Hu},\ and\ \citenamefont
  {Wu}}]{PhysRevE.82.022103}%
  \BibitemOpen
  \bibfield  {author} {\bibinfo {author} {\bibfnamefont {Q.~Y.}\ \bibnamefont
  {Hao}}, \bibinfo {author} {\bibfnamefont {R.}~\bibnamefont {Jiang}}, \bibinfo
  {author} {\bibfnamefont {M.~B.}\ \bibnamefont {Hu}}, \ and\ \bibinfo {author}
  {\bibfnamefont {Q.~S.}\ \bibnamefont {Wu}},\ }\href {\doibase
  10.1103/PhysRevE.82.022103} {\bibfield  {journal} {\bibinfo  {journal} {Phys.
  Rev. E}\ }\textbf {\bibinfo {volume} {82}},\ \bibinfo {pages} {022103}
  (\bibinfo {year} {2010})}\BibitemShut {NoStop}%
\bibitem [{\citenamefont {Hao}\ \emph {et~al.}(2016)\citenamefont {Hao},
  \citenamefont {Jiang}, \citenamefont {Hu}, \citenamefont {Jia},\ and\
  \citenamefont {Wang}}]{hao2016exponential}%
  \BibitemOpen
  \bibfield  {author} {\bibinfo {author} {\bibfnamefont {Q.~Y.}\ \bibnamefont
  {Hao}}, \bibinfo {author} {\bibfnamefont {R.}~\bibnamefont {Jiang}}, \bibinfo
  {author} {\bibfnamefont {M.-B.}\ \bibnamefont {Hu}}, \bibinfo {author}
  {\bibfnamefont {B.}~\bibnamefont {Jia}}, \ and\ \bibinfo {author}
  {\bibfnamefont {W.-X.}\ \bibnamefont {Wang}},\ }\href@noop {} {\bibfield
  {journal} {\bibinfo  {journal} {Sci. Rep.}\ }\textbf {\bibinfo {volume}
  {6}},\ \bibinfo {pages} {19652} (\bibinfo {year} {2016})}\BibitemShut
  {NoStop}%
\bibitem [{\citenamefont {Lin}\ \emph {et~al.}(2011)\citenamefont {Lin},
  \citenamefont {Steinberg},\ and\ \citenamefont
  {Ashwin}}]{lin2011bidirectional}%
  \BibitemOpen
  \bibfield  {author} {\bibinfo {author} {\bibfnamefont {C.}~\bibnamefont
  {Lin}}, \bibinfo {author} {\bibfnamefont {G.}~\bibnamefont {Steinberg}}, \
  and\ \bibinfo {author} {\bibfnamefont {P.}~\bibnamefont {Ashwin}},\
  }\href@noop {} {\bibfield  {journal} {\bibinfo  {journal} {J. Stat. Mech.
  Theory}\ }\textbf {\bibinfo {volume} {2011}},\ \bibinfo {pages} {P09027}
  (\bibinfo {year} {2011})}\BibitemShut {NoStop}%
\bibitem [{\citenamefont {Jiang}\ \emph {et~al.}(2009)\citenamefont {Jiang},
  \citenamefont {Nishinari}, \citenamefont {Hu}, \citenamefont {Wu},\ and\
  \citenamefont {Wu}}]{jiang2009phase}%
  \BibitemOpen
  \bibfield  {author} {\bibinfo {author} {\bibfnamefont {R.}~\bibnamefont
  {Jiang}}, \bibinfo {author} {\bibfnamefont {K.}~\bibnamefont {Nishinari}},
  \bibinfo {author} {\bibfnamefont {M.~B.}\ \bibnamefont {Hu}}, \bibinfo
  {author} {\bibfnamefont {Y.~H.}\ \bibnamefont {Wu}}, \ and\ \bibinfo {author}
  {\bibfnamefont {Q.~S.}\ \bibnamefont {Wu}},\ }\href@noop {} {\bibfield
  {journal} {\bibinfo  {journal} {J. Stat. Phys.}\ }\textbf {\bibinfo {volume}
  {136}},\ \bibinfo {pages} {73} (\bibinfo {year} {2009})}\BibitemShut
  {NoStop}%
\bibitem [{\citenamefont {Hao}\ \emph {et~al.}(2018)\citenamefont {Hao},
  \citenamefont {Jiang}, \citenamefont {Wu}, \citenamefont {Guo}, \citenamefont
  {Liu},\ and\ \citenamefont {Zhang}}]{hao2018theoretical}%
  \BibitemOpen
  \bibfield  {author} {\bibinfo {author} {\bibfnamefont {Q.~Y.}\ \bibnamefont
  {Hao}}, \bibinfo {author} {\bibfnamefont {R.}~\bibnamefont {Jiang}}, \bibinfo
  {author} {\bibfnamefont {C.~Y.}\ \bibnamefont {Wu}}, \bibinfo {author}
  {\bibfnamefont {N.}~\bibnamefont {Guo}}, \bibinfo {author} {\bibfnamefont
  {B.~B.}\ \bibnamefont {Liu}}, \ and\ \bibinfo {author} {\bibfnamefont
  {Y.}~\bibnamefont {Zhang}},\ }\href@noop {} {\bibfield  {journal} {\bibinfo
  {journal} {Phys. Rev. E}\ }\textbf {\bibinfo {volume} {98}},\ \bibinfo
  {pages} {062111} (\bibinfo {year} {2018})}\BibitemShut {NoStop}%
\bibitem [{\citenamefont {Hao}\ \emph {et~al.}(2019)\citenamefont {Hao},
  \citenamefont {Jiang}, \citenamefont {Hu}, \citenamefont {Zhang},
  \citenamefont {Wu},\ and\ \citenamefont {Guo}}]{PhysRevE.100.032133}%
  \BibitemOpen
  \bibfield  {author} {\bibinfo {author} {\bibfnamefont {Q.~Y.}\ \bibnamefont
  {Hao}}, \bibinfo {author} {\bibfnamefont {R.}~\bibnamefont {Jiang}}, \bibinfo
  {author} {\bibfnamefont {M.~B.}\ \bibnamefont {Hu}}, \bibinfo {author}
  {\bibfnamefont {Y.}~\bibnamefont {Zhang}}, \bibinfo {author} {\bibfnamefont
  {C.~Y.}\ \bibnamefont {Wu}}, \ and\ \bibinfo {author} {\bibfnamefont
  {N.}~\bibnamefont {Guo}},\ }\href {\doibase 10.1103/PhysRevE.100.032133}
  {\bibfield  {journal} {\bibinfo  {journal} {Phys. Rev. E}\ }\textbf {\bibinfo
  {volume} {100}},\ \bibinfo {pages} {032133} (\bibinfo {year}
  {2019})}\BibitemShut {NoStop}%
\bibitem [{\citenamefont {Ezaki}\ and\ \citenamefont
  {Nishinari}(2011)}]{ezaki2011positive}%
  \BibitemOpen
  \bibfield  {author} {\bibinfo {author} {\bibfnamefont {T.}~\bibnamefont
  {Ezaki}}\ and\ \bibinfo {author} {\bibfnamefont {K.}~\bibnamefont
  {Nishinari}},\ }\href@noop {} {\bibfield  {journal} {\bibinfo  {journal}
  {Phys. Rev. E}\ }\textbf {\bibinfo {volume} {84}},\ \bibinfo {pages} {061149}
  (\bibinfo {year} {2011})}\BibitemShut {NoStop}%
\bibitem [{\citenamefont {Tsuzuki}\ \emph
  {et~al.}(2018{\natexlab{a}})\citenamefont {Tsuzuki}, \citenamefont
  {Yanagisawa},\ and\ \citenamefont {Nishinari}}]{tsuzuki2018effect}%
  \BibitemOpen
  \bibfield  {author} {\bibinfo {author} {\bibfnamefont {S.}~\bibnamefont
  {Tsuzuki}}, \bibinfo {author} {\bibfnamefont {D.}~\bibnamefont {Yanagisawa}},
  \ and\ \bibinfo {author} {\bibfnamefont {K.}~\bibnamefont {Nishinari}},\
  }\href@noop {} {\bibfield  {journal} {\bibinfo  {journal} {Phys. Rev. E}\
  }\textbf {\bibinfo {volume} {97}},\ \bibinfo {pages} {042117} (\bibinfo
  {year} {2018}{\natexlab{a}})}\BibitemShut {NoStop}%
\bibitem [{\citenamefont {Tsuzuki}\ \emph
  {et~al.}(2018{\natexlab{b}})\citenamefont {Tsuzuki}, \citenamefont
  {Yanagisawa},\ and\ \citenamefont {Nishinari}}]{PhysRevE.98.042102}%
  \BibitemOpen
  \bibfield  {author} {\bibinfo {author} {\bibfnamefont {S.}~\bibnamefont
  {Tsuzuki}}, \bibinfo {author} {\bibfnamefont {D.}~\bibnamefont {Yanagisawa}},
  \ and\ \bibinfo {author} {\bibfnamefont {K.}~\bibnamefont {Nishinari}},\
  }\href {\doibase 10.1103/PhysRevE.98.042102} {\bibfield  {journal} {\bibinfo
  {journal} {Phys. Rev. E}\ }\textbf {\bibinfo {volume} {98}},\ \bibinfo
  {pages} {042102} (\bibinfo {year} {2018}{\natexlab{b}})}\BibitemShut
  {NoStop}%
\bibitem [{\citenamefont {Parmeggiani}\ \emph {et~al.}(2003)\citenamefont
  {Parmeggiani}, \citenamefont {Franosch},\ and\ \citenamefont
  {Frey}}]{PhysRevLett.90.086601}%
  \BibitemOpen
  \bibfield  {author} {\bibinfo {author} {\bibfnamefont {A.}~\bibnamefont
  {Parmeggiani}}, \bibinfo {author} {\bibfnamefont {T.}~\bibnamefont
  {Franosch}}, \ and\ \bibinfo {author} {\bibfnamefont {E.}~\bibnamefont
  {Frey}},\ }\href {\doibase 10.1103/PhysRevLett.90.086601} {\bibfield
  {journal} {\bibinfo  {journal} {Phys. Rev. Lett.}\ }\textbf {\bibinfo
  {volume} {90}},\ \bibinfo {pages} {086601} (\bibinfo {year}
  {2003})}\BibitemShut {NoStop}%
\bibitem [{\citenamefont {Evans}\ \emph {et~al.}(2003)\citenamefont {Evans},
  \citenamefont {Juh\'asz},\ and\ \citenamefont {Santen}}]{PhysRevE.68.026117}%
  \BibitemOpen
  \bibfield  {author} {\bibinfo {author} {\bibfnamefont {M.~R.}\ \bibnamefont
  {Evans}}, \bibinfo {author} {\bibfnamefont {R.}~\bibnamefont {Juh\'asz}}, \
  and\ \bibinfo {author} {\bibfnamefont {L.}~\bibnamefont {Santen}},\ }\href
  {\doibase 10.1103/PhysRevE.68.026117} {\bibfield  {journal} {\bibinfo
  {journal} {Phys. Rev. E}\ }\textbf {\bibinfo {volume} {68}},\ \bibinfo
  {pages} {026117} (\bibinfo {year} {2003})}\BibitemShut {NoStop}%
\bibitem [{\citenamefont {Parmeggiani}\ \emph {et~al.}(2004)\citenamefont
  {Parmeggiani}, \citenamefont {Franosch},\ and\ \citenamefont
  {Frey}}]{PhysRevE.70.046101}%
  \BibitemOpen
  \bibfield  {author} {\bibinfo {author} {\bibfnamefont {A.}~\bibnamefont
  {Parmeggiani}}, \bibinfo {author} {\bibfnamefont {T.}~\bibnamefont
  {Franosch}}, \ and\ \bibinfo {author} {\bibfnamefont {E.}~\bibnamefont
  {Frey}},\ }\href {\doibase 10.1103/PhysRevE.70.046101} {\bibfield  {journal}
  {\bibinfo  {journal} {Phys. Rev. E}\ }\textbf {\bibinfo {volume} {70}},\
  \bibinfo {pages} {046101} (\bibinfo {year} {2004})}\BibitemShut {NoStop}%
\bibitem [{\citenamefont {Wang}\ \emph {et~al.}(2008)\citenamefont {Wang},
  \citenamefont {Liu},\ and\ \citenamefont {Jiang}}]{WANG2008457}%
  \BibitemOpen
  \bibfield  {author} {\bibinfo {author} {\bibfnamefont {R.}~\bibnamefont
  {Wang}}, \bibinfo {author} {\bibfnamefont {M.}~\bibnamefont {Liu}}, \ and\
  \bibinfo {author} {\bibfnamefont {R.}~\bibnamefont {Jiang}},\ }\href
  {\doibase https://doi.org/10.1016/j.physa.2007.09.042} {\bibfield  {journal}
  {\bibinfo  {journal} {Physica A}\ }\textbf {\bibinfo {volume} {387}},\
  \bibinfo {pages} {457 } (\bibinfo {year} {2008})}\BibitemShut {NoStop}%
\bibitem [{\citenamefont {Dhiman}\ and\ \citenamefont
  {Gupta}(2014)}]{dhiman2014two}%
  \BibitemOpen
  \bibfield  {author} {\bibinfo {author} {\bibfnamefont {I.}~\bibnamefont
  {Dhiman}}\ and\ \bibinfo {author} {\bibfnamefont {A.~K.}\ \bibnamefont
  {Gupta}},\ }\href@noop {} {\bibfield  {journal} {\bibinfo  {journal}
  {Europhys. Lett.}\ }\textbf {\bibinfo {volume} {107}},\ \bibinfo {pages}
  {20007} (\bibinfo {year} {2014})}\BibitemShut {NoStop}%
\bibitem [{\citenamefont {Sharma}\ and\ \citenamefont
  {Gupta}(2017)}]{sharma2017phase}%
  \BibitemOpen
  \bibfield  {author} {\bibinfo {author} {\bibfnamefont {N.}~\bibnamefont
  {Sharma}}\ and\ \bibinfo {author} {\bibfnamefont {A.}~\bibnamefont {Gupta}},\
  }\href@noop {} {\bibfield  {journal} {\bibinfo  {journal} {J. Stat. Mech.
  Theory}\ }\textbf {\bibinfo {volume} {2017}},\ \bibinfo {pages} {043211}
  (\bibinfo {year} {2017})}\BibitemShut {NoStop}%
\bibitem [{\citenamefont {Garg}\ and\ \citenamefont
  {Dhiman}(2019)}]{GARG2019123356}%
  \BibitemOpen
  \bibfield  {author} {\bibinfo {author} {\bibfnamefont {S.}~\bibnamefont
  {Garg}}\ and\ \bibinfo {author} {\bibfnamefont {I.}~\bibnamefont {Dhiman}},\
  }\href {\doibase https://doi.org/10.1016/j.physa.2019.123356} {\bibfield
  {journal} {\bibinfo  {journal} {Physica A}\ ,\ \bibinfo {pages} {123356}}
  (\bibinfo {year} {2019})}\BibitemShut {NoStop}%
\bibitem [{\citenamefont {Vuijk}\ \emph {et~al.}(2015)\citenamefont {Vuijk},
  \citenamefont {Rens}, \citenamefont {Vahabi}, \citenamefont {MacKintosh},\
  and\ \citenamefont {Sharma}}]{vuijk2015driven}%
  \BibitemOpen
  \bibfield  {author} {\bibinfo {author} {\bibfnamefont {H.}~\bibnamefont
  {Vuijk}}, \bibinfo {author} {\bibfnamefont {R.}~\bibnamefont {Rens}},
  \bibinfo {author} {\bibfnamefont {M.}~\bibnamefont {Vahabi}}, \bibinfo
  {author} {\bibfnamefont {F.}~\bibnamefont {MacKintosh}}, \ and\ \bibinfo
  {author} {\bibfnamefont {A.}~\bibnamefont {Sharma}},\ }\href@noop {}
  {\bibfield  {journal} {\bibinfo  {journal} {Physical Review E}\ }\textbf
  {\bibinfo {volume} {91}},\ \bibinfo {pages} {032143} (\bibinfo {year}
  {2015})}\BibitemShut {NoStop}%
\bibitem [{\citenamefont {Ichiki}\ \emph
  {et~al.}(2016{\natexlab{a}})\citenamefont {Ichiki}, \citenamefont {Sato},\
  and\ \citenamefont {Nishinari}}]{ichiki2016totally}%
  \BibitemOpen
  \bibfield  {author} {\bibinfo {author} {\bibfnamefont {S.}~\bibnamefont
  {Ichiki}}, \bibinfo {author} {\bibfnamefont {J.}~\bibnamefont {Sato}}, \ and\
  \bibinfo {author} {\bibfnamefont {K.}~\bibnamefont {Nishinari}},\ }\href@noop
  {} {\bibfield  {journal} {\bibinfo  {journal} {Eur. Phys. J. B}\ }\textbf
  {\bibinfo {volume} {89}},\ \bibinfo {pages} {135} (\bibinfo {year}
  {2016}{\natexlab{a}})}\BibitemShut {NoStop}%
\bibitem [{\citenamefont {Ichiki}\ \emph
  {et~al.}(2016{\natexlab{b}})\citenamefont {Ichiki}, \citenamefont {Sato},\
  and\ \citenamefont {Nishinari}}]{ichiki2016totally2}%
  \BibitemOpen
  \bibfield  {author} {\bibinfo {author} {\bibfnamefont {S.}~\bibnamefont
  {Ichiki}}, \bibinfo {author} {\bibfnamefont {J.}~\bibnamefont {Sato}}, \ and\
  \bibinfo {author} {\bibfnamefont {K.}~\bibnamefont {Nishinari}},\ }\href@noop
  {} {\bibfield  {journal} {\bibinfo  {journal} {J. Phys. Soc. Jpn.}\ }\textbf
  {\bibinfo {volume} {85}},\ \bibinfo {pages} {044001} (\bibinfo {year}
  {2016}{\natexlab{b}})}\BibitemShut {NoStop}%
\bibitem [{\citenamefont {Yanagisawa}\ and\ \citenamefont
  {Ichiki}(2016)}]{yanagisawa2016totally}%
  \BibitemOpen
  \bibfield  {author} {\bibinfo {author} {\bibfnamefont {D.}~\bibnamefont
  {Yanagisawa}}\ and\ \bibinfo {author} {\bibfnamefont {S.}~\bibnamefont
  {Ichiki}},\ }in\ \href@noop {} {\emph {\bibinfo {booktitle} {International
  Conference on Cellular Automata}}}\ (\bibinfo {organization} {Springer},\
  \bibinfo {year} {2016})\ pp.\ \bibinfo {pages} {405--412}\BibitemShut
  {NoStop}%
\bibitem [{\citenamefont {Midha}\ \emph {et~al.}(2018)\citenamefont {Midha},
  \citenamefont {Kolomeisky},\ and\ \citenamefont
  {Gupta}}]{PhysRevE.98.042119}%
  \BibitemOpen
  \bibfield  {author} {\bibinfo {author} {\bibfnamefont {T.}~\bibnamefont
  {Midha}}, \bibinfo {author} {\bibfnamefont {A.~B.}\ \bibnamefont
  {Kolomeisky}}, \ and\ \bibinfo {author} {\bibfnamefont {A.~K.}\ \bibnamefont
  {Gupta}},\ }\href {\doibase 10.1103/PhysRevE.98.042119} {\bibfield  {journal}
  {\bibinfo  {journal} {Phys. Rev. E}\ }\textbf {\bibinfo {volume} {98}},\
  \bibinfo {pages} {042119} (\bibinfo {year} {2018})}\BibitemShut {NoStop}%
\bibitem [{\citenamefont {Nishinari}\ \emph {et~al.}(2005)\citenamefont
  {Nishinari}, \citenamefont {Okada}, \citenamefont {Schadschneider},\ and\
  \citenamefont {Chowdhury}}]{nishinari2005intracellular}%
  \BibitemOpen
  \bibfield  {author} {\bibinfo {author} {\bibfnamefont {K.}~\bibnamefont
  {Nishinari}}, \bibinfo {author} {\bibfnamefont {Y.}~\bibnamefont {Okada}},
  \bibinfo {author} {\bibfnamefont {A.}~\bibnamefont {Schadschneider}}, \ and\
  \bibinfo {author} {\bibfnamefont {D.}~\bibnamefont {Chowdhury}},\ }\href@noop
  {} {\bibfield  {journal} {\bibinfo  {journal} {Phys. Rev. Lett.}\ }\textbf
  {\bibinfo {volume} {95}},\ \bibinfo {pages} {118101} (\bibinfo {year}
  {2005})}\BibitemShut {NoStop}%
\bibitem [{\citenamefont {Miedema}\ \emph {et~al.}(2017)\citenamefont
  {Miedema}, \citenamefont {Kushwaha}, \citenamefont {Denisov}, \citenamefont
  {Acar}, \citenamefont {Nienhuis}, \citenamefont {Peterman},\ and\
  \citenamefont {Schall}}]{miedema2017correlation}%
  \BibitemOpen
  \bibfield  {author} {\bibinfo {author} {\bibfnamefont {D.~M.}\ \bibnamefont
  {Miedema}}, \bibinfo {author} {\bibfnamefont {V.~S.}\ \bibnamefont
  {Kushwaha}}, \bibinfo {author} {\bibfnamefont {D.~V.}\ \bibnamefont
  {Denisov}}, \bibinfo {author} {\bibfnamefont {S.}~\bibnamefont {Acar}},
  \bibinfo {author} {\bibfnamefont {B.}~\bibnamefont {Nienhuis}}, \bibinfo
  {author} {\bibfnamefont {E.~J.}\ \bibnamefont {Peterman}}, \ and\ \bibinfo
  {author} {\bibfnamefont {P.}~\bibnamefont {Schall}},\ }\href@noop {}
  {\bibfield  {journal} {\bibinfo  {journal} {Physical Review X}\ }\textbf
  {\bibinfo {volume} {7}},\ \bibinfo {pages} {041037} (\bibinfo {year}
  {2017})}\BibitemShut {NoStop}%
\bibitem [{\citenamefont {Kushwaha}\ \emph {et~al.}(2020)\citenamefont
  {Kushwaha}, \citenamefont {Acar}, \citenamefont {Miedema}, \citenamefont
  {Denisov}, \citenamefont {Schall},\ and\ \citenamefont
  {Peterman}}]{kushwaha2020crowding}%
  \BibitemOpen
  \bibfield  {author} {\bibinfo {author} {\bibfnamefont {V.~S.}\ \bibnamefont
  {Kushwaha}}, \bibinfo {author} {\bibfnamefont {S.}~\bibnamefont {Acar}},
  \bibinfo {author} {\bibfnamefont {D.~M.}\ \bibnamefont {Miedema}}, \bibinfo
  {author} {\bibfnamefont {D.~V.}\ \bibnamefont {Denisov}}, \bibinfo {author}
  {\bibfnamefont {P.}~\bibnamefont {Schall}}, \ and\ \bibinfo {author}
  {\bibfnamefont {E.~J.}\ \bibnamefont {Peterman}},\ }\href@noop {} {\bibfield
  {journal} {\bibinfo  {journal} {Plos one}\ }\textbf {\bibinfo {volume}
  {15}},\ \bibinfo {pages} {e0228930} (\bibinfo {year} {2020})}\BibitemShut
  {NoStop}%
\end{thebibliography}
\end{document}